\documentclass[traditabstract]{aa}  
\usepackage[varg]{txfonts}
\usepackage{graphicx}

\usepackage{natbib}
\usepackage{float}
\usepackage{hyperref}

\begin{document}

\title{Molecular gas in a system of two interacting galaxies overlapping on the line-of-sight}
\titlerunning{Molecular gas in two interacting galaxies}
\authorrunning{Halle et al}

\author{Ana\"elle Halle\inst{1}, Barbara Mazzilli Ciraulo\inst{1,2,3}, Daniel Maschmann\inst{1,4}, Anne-Laure Melchior\inst{1} and Francoise Combes\inst{1, 5}\\ 
}
\institute{LUX, Observatoire de Paris, Sorbonne Universit\'{e}, Universit\'{e} PSL, CNRS, F-75014, Paris, France\\
\email{anaelle.halle@observatoiredeparis.psl.eu}
\and
Centre for Astrophysics and Supercomputing, Swinburne University of Technology, P.O. Box 218, Hawthorn, VIC 3122, Australia
\and
ARC Centre of Excellence for All Sky Astrophysics in 3 Dimensions (ASTRO 3D), Australia
\and
Steward Observatory, University of Arizona, 933 N Cherry Ave, Tucson, AZ 85721, USA
\and
Coll\`ege de France, 11, Place Marcelin Berthelot, F-75005, Paris, France
}

\date{}

\abstract {
Galaxy interactions can disturb gas in galactic discs, compress it, excite it, and enhance star formation. An intriguing system likely consisting of two interacting galaxies overlapping on the line-of-sight was previously studied through ionised gas observations from the integral-field spectrograph Mapping Nearby Galaxies at APO (MaNGA). A decomposition into two components using MaNGA spectra, together with a multi-wavelength study, allowed to characterise the system as a minor-merger with interaction-induced star-formation, and maybe AGN activity. We use new interferometric observations of the CO(1-0) gas of this system from the NOrthern Extended Millimeter Array (NOEMA) to confront and combine the spatially-resolved ionised and molecular gas observations. Mock NOEMA and MaNGA data are computed from simulated systems of two discs and compared to the observations. The NOEMA observations of the molecular gas, dynamically colder than the ionised gas, help to precise the configuration of the system, which we revisit as being a major merger. Combination of ionised and molecular gas data allow us to study the star-formation efficiency of the system.}


\maketitle

\section{Introduction}\label{sect:introduction} 

Galaxy interactions are sites of possible gas compression, enhancement of star formation, and triggering of Active Galactic Nuclei (AGN) activity. The molecular gas is a particularly interesting tracer to study the dynamics of such systems, as it is dynamically cold. It can be observed through the emission of the CO molecule in a large range of redshifts. 

Combining resolved molecular gas and star-formation rate (SFR) observations, Kennicutt-Schmidt relations \citep{1959ApJ...129..243S, 1998ARA&A..36..189K} have been obtained in local galaxies \citep{2008AJ....136.2782L, 2008AJ....136.2846B}. Similar studies at higher redshift have also been performed, as by \citet{freundlich13} around redshift 1.2. For example, it is possible to estimate the SFR from ionised gas, and in particular
the H${\alpha}$ emission of recombining hydrogen. Studies such as ALMaQUEST \citep[][]{lin19} have used ionised-gas and molecular-gas resolved observations in low-redshift galaxies to derive star formation efficiency and departure from the main sequence of star-forming galaxies, relating SFR and stellar mass. 

This work focuses on J221024.49+114247.0, a $z=0.09228$ (according to the NASA Sloan Atlas catalogue) system which was identified as showing double-peaked emission-lines of ionised gas in the $3''$ SDSS fibre by \citet{maschmann20}. Double-peaked emission-lines have a variety of possible origins including disc rotation, gas outflows, and galaxy mergers \citep{maschmann23}. In the case of J221024.49+114247.0, a detailed study was performed in \citet{mazzilli21}, combining integral-field-unit (IFU) MaNGA observations \citep{2015ApJ...798....7B} of the ionised gas (with the identifier MaNGA-ID 1-114955, or, in Plate-IFU, 7977-12701), IRAM 30~m observations of the CO(1-0) and CO(2-1) emission, and VLA radio observations. Through the decomposition of the spatially resolved MaNGA observations of the ionised gas, it is possible to isolate two different kinematic components corresponding to two counter-rotating galactic discs overlapping on the line-of-sight. 

New interferometric observations of the molecular gas were performed by the NOrthern Extended Millimeter Array (NOEMA), in the CO(1-0) line, yielding the spatially-resolved molecular content of the system, to be compared with the spatially-resolved ionised-gas data.   

After a presentation of the system of interest and a summary of related previous works in Sec.~\ref{sec:pres}, we study its molecular gas content with NOEMA and the relationship between this gas and SFR obtained through H$\alpha$ in Sec.~\ref{sec:mol-ion}. We then model the system kinematically with two discs and recover the contribution of each galaxy to the molecular gas content and to the emission-lines of ionised gas in Sec.~\ref{sec:model}; we discuss our results in Sec.~\ref{sec:discu}. 

Throughout this work, we assume a flat $\Lambda$CDM cosmology with $\Omega_M=0.3$, $\Omega_{\Lambda} = 0.7$ and $H_0=70$~km/s/Mpc.

\section{J221024.49+114247.0}
\label{sec:pres}
\begin{figure*}
    \centering
    \centering
    \begin{tabular}{cc}
    \includegraphics[height = 0.42\linewidth]{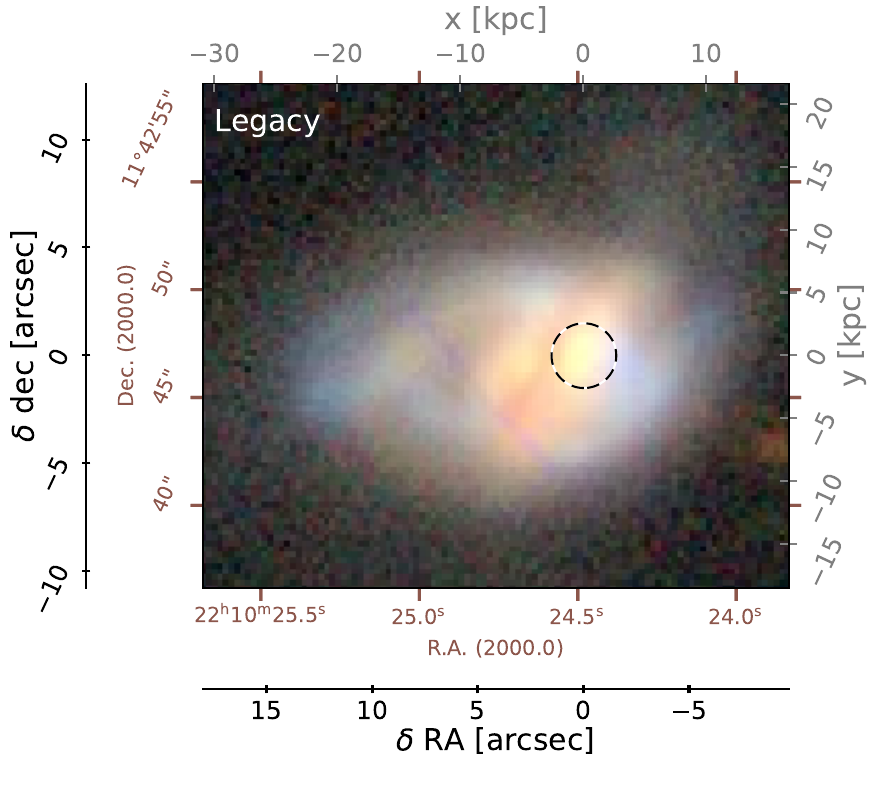}
         &
    \includegraphics[height = 0.42\linewidth]{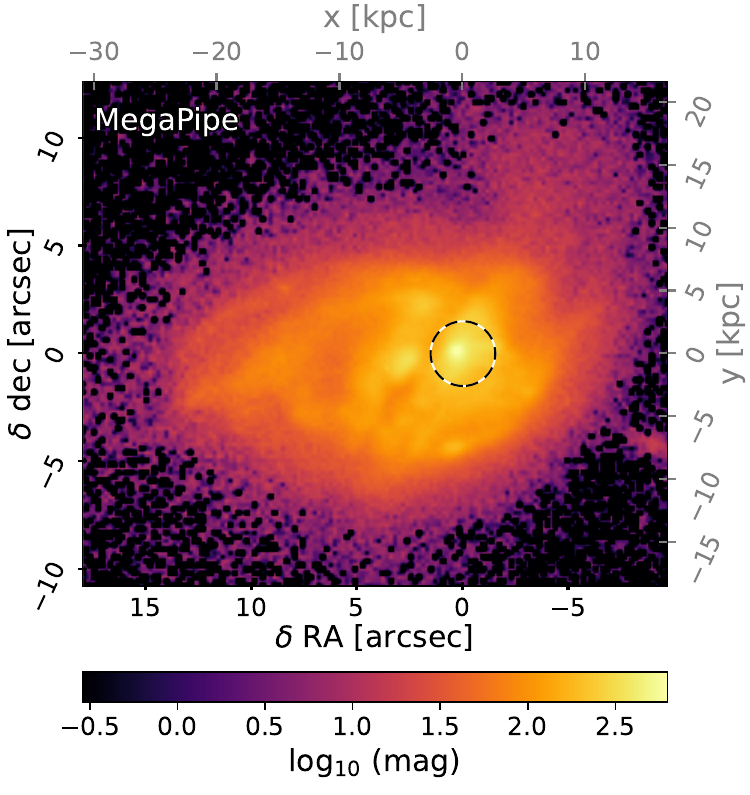} \\
    \end{tabular}
    \caption{Legacy composite image (left) and $\log_{10}$ of MegaPipe image (right), from MegaCam on CFHT. The 3$''$ SDSS fibre is over-plotted in a dashed black circle.}
    \label{fig:legacy-megapipe}
\end{figure*}

The system J221024.49+114247.0 appears in optical and near-infrared bands as a perturbed galaxy. Figure~\ref{fig:legacy-megapipe} shows a composite Legacy image \citep{2019AJ....157..168D} and a CFHT MegaPipe \citep{2008PASP..120..212G} $r$-band image. It has a redshift of $z=0.09228$ according to the NASA Sloan Atlas catalogue. The SDSS fibre is represented as a black dashed circle in Fig.~\ref{fig:legacy-megapipe} and other maps of this work. \citet{maschmann20} identified double-peaked emission-lines in this fibre. According to the spectral-energy distribution (SED) fitting of \citet{2016ApJS..227....2S}, the stellar mass is 1.6~$10^{10}$~M$_{\odot}$ and the SFR is 14~M$_{\odot}$/yr.  

As detailed in \citet{mazzilli21}, the system exhibits spiral features, possible tidal features (especially on the North-West, as best seen on the CFHT MegaPipe image of Fig.~\ref{fig:legacy-megapipe}), and what appears to be a dust lane from South-East to North-West. Coordinates relative to the centre of the SDSS fibre are shown, in arcseconds, and in kiloparsecs \footnote{Transformations between different world coordinate systems and a significant amount of plotting in the present work are performed with \texttt{astropy} \citep{astropy:2013}.}. 

Using MaNGA observations, \citet{mazzilli21} concluded that the system is a minor merger, of mass ratio 1:9, of two disc galaxies overlapping on the line-of-sight: a ``main galaxy'', extending on the whole spatial extent of the observed system, and a ``secondary galaxy'', smaller and likely aligned with the dust lane visible on the legacy composite image of Fig.~\ref{fig:legacy-megapipe}. Their study was based on decompositions of both stellar and gas emission-line MaNGA data into two components in each spaxel. For gas emission-lines, both single and double Gaussian fits of emission-lines were performed in each spaxel, with limits on the variation of fitted parameters between neighbouring spaxels, and an $F$-test determined the best of single and double Gaussian fits for each spaxel. The mass ratio of 1:9 was estimated from the rotation velocities obtained from the histogram of stellar or gas velocities corrected from the estimated inclinations, and from the estimated spatial extents of the two discs. The spectral index obtained from radio data was not conclusive regarding possible AGN activity. This system is a member of the 69 double-peaked galaxies sample of \citet{mazzilli24}, where it is labelled G61. In \citet{mazzilli24}, G61 is shown to belong to the blue cloud with an apparent NUV-$r$ magnitude of 2.4 and an intrinsic corrected NUV-$r$ magnitude (corrected for dust attenuation) of 2. Its $r$-band absolute magnitude $M_r$ is of -22.7. It is also shown to belong to the star-forming sequence of the intrinsic NUV-$r$ magnitude versus specific SFR diagram (see Fig.10 of \citealt{mazzilli24}). Its stellar index $D_n$~4000 is below 1.5 in most of the galaxy, indicating young stellar populations (see Fig.~G2 of \citealt{mazzilli24}). 

\citet{mazzilli21} used the MaNGA DR16 data to perform emission-line fitting for this system. They used the VOR10-GAU-MILESHC model cube from the MaNGA Data Analysis Pipeline (DAP) \citep{2019AJ....158..231W}, with the same Voronoi binning for stellar and gaseous components, because they also performed a study of the stellar component with \texttt{Nburst} \citep{nburst_a, nburst_b}. MaNGA DR17 data have become available, with a new flux calibration and new stellar and emission-line fittings. In the present work, we focus on the study of emission-lines (which we compare to the molecular gas data) and thus use the DR17 HYB10-MILESHC-MASTARSSP cube, with a hybrid binning scheme (Voronoi binning for stars but not for gas), recommended in the case of study of the sole emission-lines. After removing the stellar contribution using the DAP model cube and correcting for Galactic extinction, we perform the same fitting procedure on the emission-lines as in \citet{mazzilli21}. These fits are used to produce the maps and diagrams of the next section (Sec.~\ref{sec:mol-ion}) and App.~\ref{app:manga}.

\section{Molecular gas content and star formation}
\label{sec:mol-ion}
We study the molecular gas content of the system based on NOEMA CO(1-0) observations. We use MaNGA observations to  compute metallicity and derive the conversion factor from CO(1-0) emission to molecular gas mass. From the surface densities of both molecular gas and SFR from MaNGA, we compute the star-formation efficiency. 

\subsection{NOEMA observations of CO(1-0)}
\label{sec-noemaobs}
\begin{figure*}
    \centering
    \includegraphics[width = \linewidth]{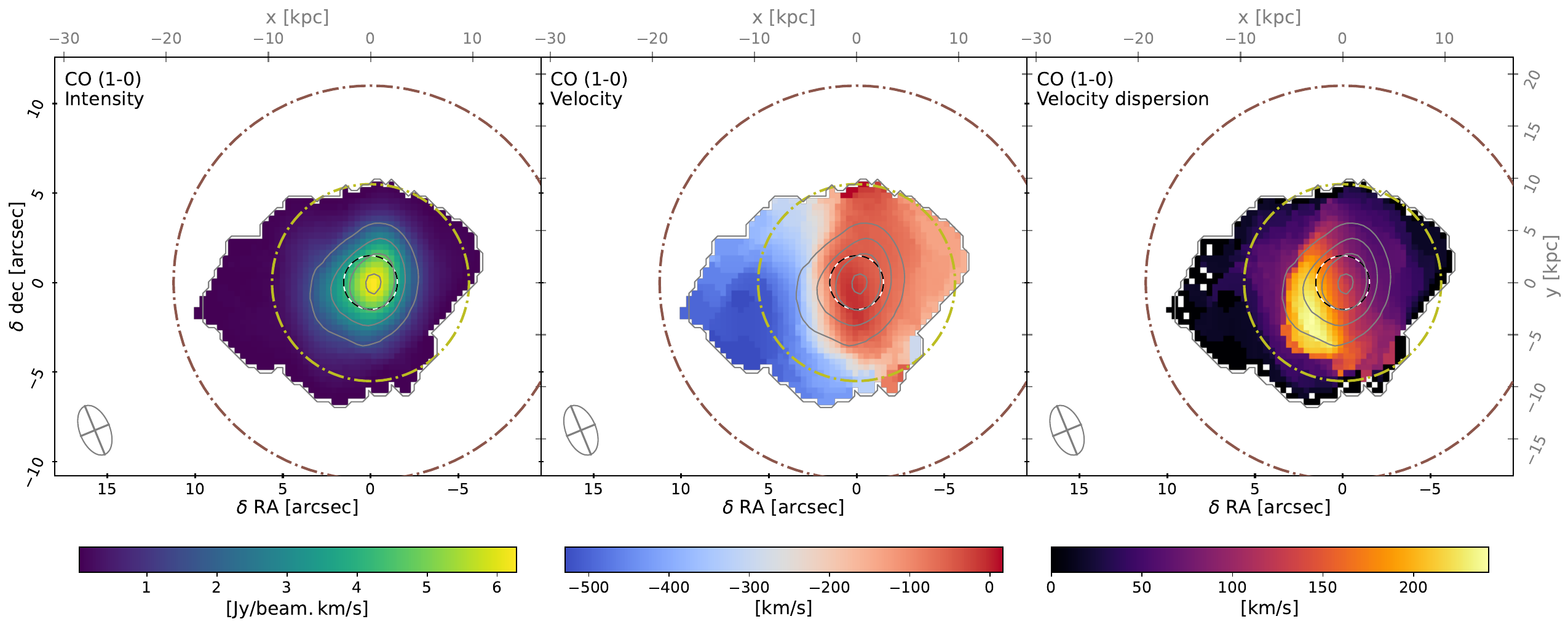}
\caption{Intensity, velocity and velocity dispersion obtained for the CO(1-0) NOEMA observations. A flux density threshold of 1~mJy/beam is applied. We over-plot linearly-spaced contours of the CO(1-0) intensity on each map. The NOEMA elliptical Gaussian synthesised beam is represented in the bottom-left corners of the panels. The IRAM~30~m CO(1-0) and CO(2-1) beams are shown in brown and yellow, respectively.   The 3$''$ SDSS fibre is over-plotted in a dashed black circle.}
    \label{fig:noema-moments}
\end{figure*}

\begin{figure}
    \centering
    \includegraphics[width = \linewidth]{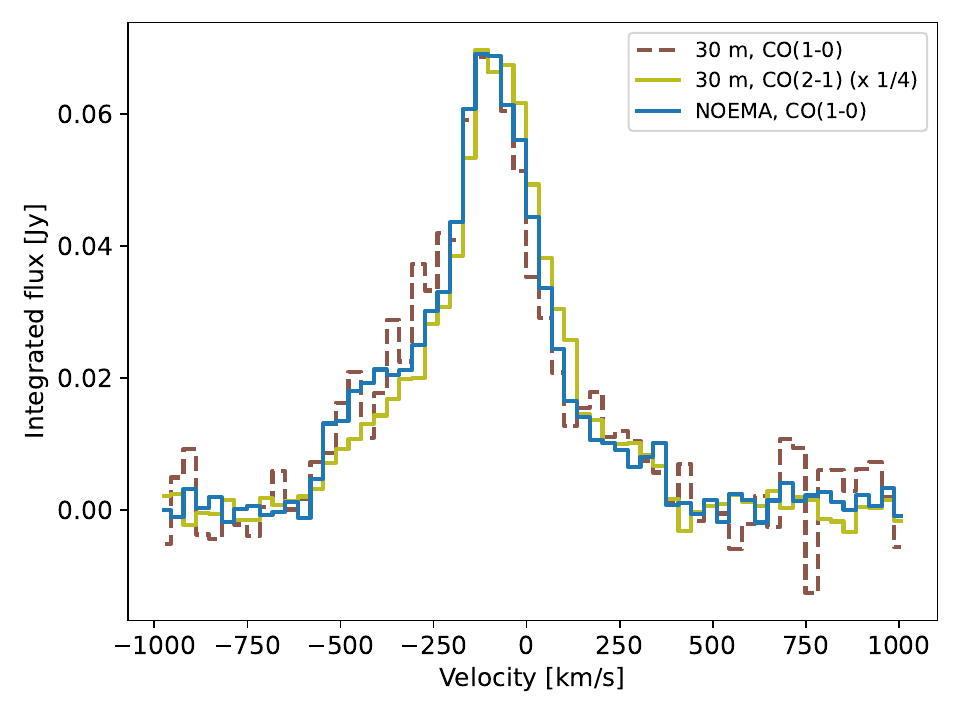}
    \caption{NOEMA CO(1-0) spectrum stacked over a 22 arcsec aperture, compared to the IRAM 30~m spectra, 
    at the same velocity resolution.}
    \label{fig:co-spectra}
\end{figure}

\begin{figure*}[h]
    \centering
    \includegraphics[width = \linewidth]{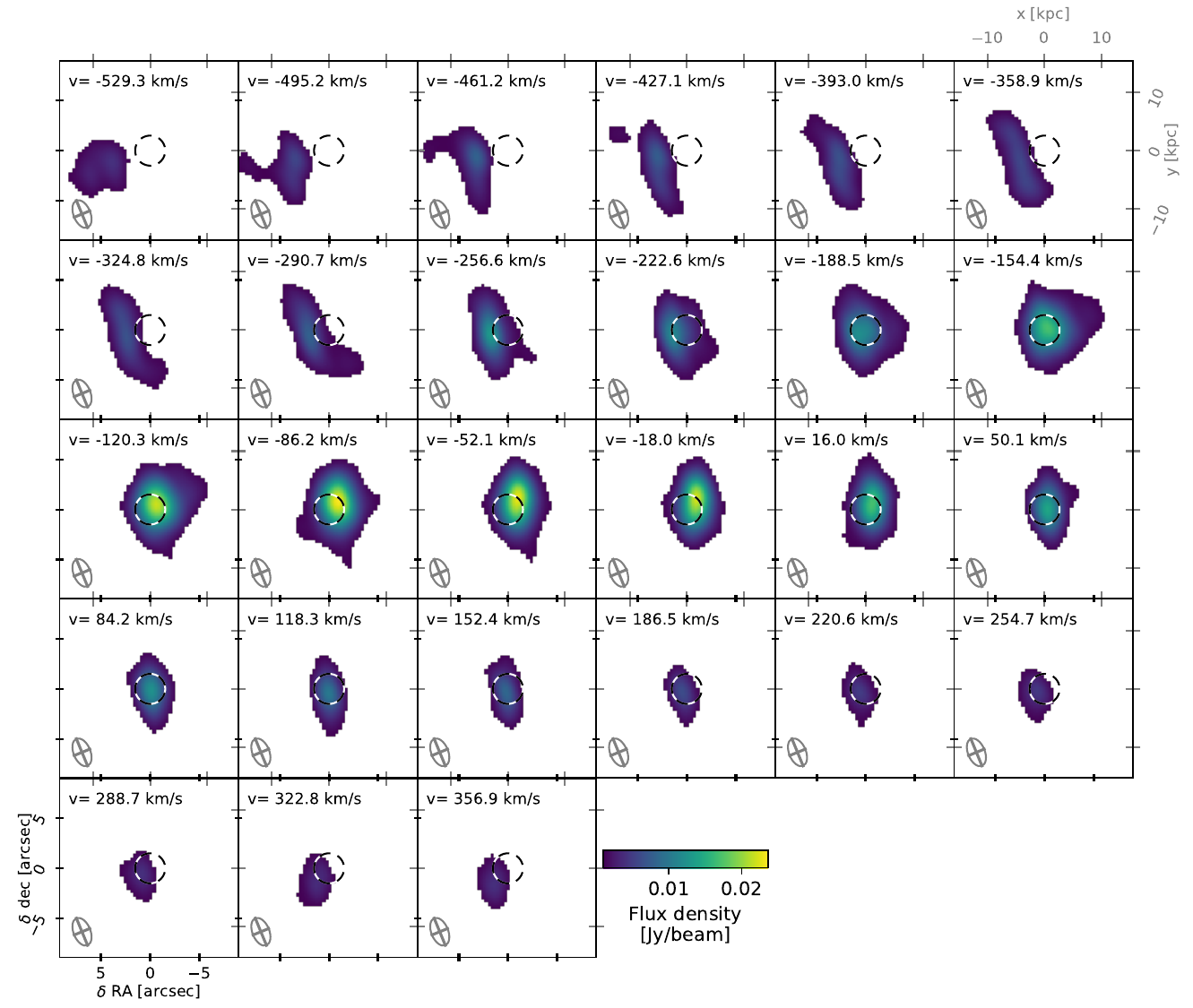}
    \caption{Velocity channel maps. The 3$''$ SDSS fibre is over-plotted in a dashed black circle.} 
    \label{fig:channels}
\end{figure*}

Single-dish observations of J221024.49+114247.0 in CO(1-0) and CO(2-1) were performed at the IRAM 30~m telescope and presented in \citet{mazzilli21}. The source was then observed in CO(1-0) by NOEMA (project S21BQ001, P.I. Barbara Mazzilli Ciraulo), at a rest frequency of 105.533~GHz, during three time-slots in 2021: 2.1 hours on the 21$^{\rm st}$ of July in configuration 10D of the array, 1.1 hour on the 25$^{\rm th}$ of July in configuration 8D+N01, and 2.6 hours on the 25$^{\rm th}$ of November in configuration 10C. Data calibration and image deconvolution were performed with the \texttt{GILDAS}~\footnote{\href{https://www.iram.fr/IRAMFR/GILDAS}{https://www.iram.fr/IRAMFR/GILDAS}} software, using natural weighting and a Hogbom cleaning on the whole field of view of $95.5''  \times 95.5''$. The elliptical Gaussian synthesised beam has axes of full widths at mid-heights (FWHMs) of $2.92''$ and $1.68''$, with a $22.4^{\circ}$ position-angle. Table.~\ref{tab:obs-resolution} gathers spatial resolution details with conversions into kpc (with similar details for MaNGA observations). NOEMA velocity channels are stacked 6 by 6 to increase the signal-to-noise ratio, resulting in a velocity resolution of 34.1~km/s.

\begin{table}[h]
 \caption{Spatial resolution of NOEMA and MaNGA observations.} 
    \centering
    \begin{tabular}{ccc}
         \hline
         \hline
          Observation & Pixel width &  Modelled PSF/beam \\
         \hline
         NOEMA & 0.32'' & Elliptical Gaussian \\
         & (0.55~kpc) & with PA=$22.4^{\circ}$, \\
         & & FWHMs = $2.92''$ and $1.68''$ \\
         & & (5.01 and 2.88~kpc) \\
         \hline
         MaNGA &  0.5''  & Circular Gaussian \\
         & (0.86~kpc) & with FWHM = 2.6''  \\
         & & (4.46~kpc) in $g$ to $z$ bands\\
         \hline
    \end{tabular}
    \label{tab:obs-resolution}
\end{table}

Maps of the CO(1-0) line intensity, line-of-sight velocity and velocity dispersion, obtained through summations on velocity channels, are shown on Fig~\ref{fig:noema-moments}\footnote{The first two maps are the usual moments of observational studies of molecular gas and the third one should be multiplied by $\sqrt{8 \log 2}$ to obtain the more usually defined second moment.}. A threshold in flux density was applied in summations, to limit the noise level\footnote{The threshold value was lowered in combination with a cleaning procedure removing small groups of voxels remaining after cutting below a threshold.}. Velocity channel maps, with the same threshold, are shown in Fig.~\ref{fig:channels}. When velocity increases, the peak flux density shifts from the South-West (bottom right of each panel) to the North-East (top left of each panel) at $v \simeq 0$, and to the South-West again. A 3D view of the NOEMA cube, using the same threshold, is shown in Fig.~\ref{fig:noema-cube}, with a link to the corresponding 3D interactive visualisation\footnote{Several position-velocity diagrams of this complex NOEMA cube are shown in the analyses of Sec.~\ref{sec:model}, either after integration over one spatial direction or for thin slices but after rotation of the NOEMA cube using position-angles of modelled discs.}.

The CO(1-0) NOEMA spectrum, gathered over the 30~m beam of 11 arcsec radius, is compared in Fig.~\ref{fig:co-spectra} with the CO(1-0) 30~m spectrum, binned to the same velocity resolution. The two spectra are very similar, meaning that the interferometer does not lose flux. In \citet{mazzilli21}, the CO(2-1) 30~m spectrum was also shown, exhibiting a slightly different shape from the CO(1-0) at low velocities. This CO(2-1) spectrum is also shown in Fig.~\ref{fig:co-spectra}, with the same velocity resolution. It is multiplied by $1/4$ for comparison of shape with the CO(1-0) spectrum. The slight depletion at low velocities is consistent with the low-velocities part of the system being outside of the CO(2-1) beam, as visible on the velocity map of Fig.\ref{fig:noema-moments}. 

\subsection{Surface density of molecular gas}

The surface density of molecular gas is obtained through the computation of the intrinsic CO luminosity and a luminosity-to-molecular-gas-mass conversion factor.

The intrinsic CO luminosity is obtained from the velocity integrated transition line flux ${\rm F_{CO(1 \rightarrow 0)}}$ by
\begin{equation}\label{eq:lprime_co}
    \frac{L'_{\rm CO(1 \rightarrow 0)}}{\rm K \, km \, s^{-1} \, pc^{2}} = \frac{3.25 \times 10^{7}}{(1 + z)} \left( \frac{F_{\rm CO(1 \rightarrow 0)}}{\rm Jy \, km \, s^{-1}} \right)  \left( \frac{\nu_{\rm rest}}{\rm GHz} \right)^{-2}  \left( \frac{D_{L}}{\rm Mpc} \right)^{2},
\end{equation}
where $\nu_{\rm rest} = 115.271$~GHz is the rest CO(1-0) line frequency and $D_L$ the luminosity distance \citep{1997ApJ...478..144S}. The total molecular gas mass is then:
\begin{equation}\label{eq:co2mh2}
    M_{\rm mol\,gas} = \alpha_{\rm CO} \, L'_{\rm CO(1 \rightarrow 0)},
\end{equation}
with $\alpha_{\rm CO}$ the CO(1-0) luminosity-to-molecular-gas-mass conversion factor, defined to include helium in the molecular mass ($M_{\rm mol\,gas} = 1.36 \,  M_{\rm H_2}$). This factor has a value empirically established in the Milky Way of $\alpha_G = 4.36 \pm 0.9$~M$_{\odot}$~${(\rm K \, km \, s^{-1} \, pc^{2}})^{-1}$, based on assumptions including a solar metallicity for the gas. 

The conversion factor $\alpha_{\rm CO}$ is a decreasing function of the metallicity of the gas, which is parametrised for example by the oxygen-abundance as ${\rm log\,Z = 12 + log(O/H)}$. We compute this gas-phase metallicity through the $O_3N_2$ index $O_3N_2={ \rm [O III] \lambda 5007/H \beta )/([N II] \lambda 6583/ H \alpha}$, using the calibration of \citet{2004MNRAS.348L..59P}: $\log \, Z = 8.73 - 0.32 \, \log  O_3N_2$. We estimate a single value of this factor for the system through stacked Gaussian emission-line fluxes given in the MaNGA DAP catalogue, for a stacking performed in an elliptical aperture of semi-major axis the effective radius in the $r$-band of the system, as obtained from a Petrosian analysis. The corresponding $O_3N_2$ metallicity is $\log Z = 8.65$. 
Maps of the fluxes of H$\alpha$, H$\beta$, $[$OIII$]\lambda 5007$ and $[$NII$] \lambda 6583$ with a minimum signal-to-noise ratio (SNR) of 3 are shown in Fig.~\ref{fig:manga-maps}, with the gas-phase metallicity map from the $O_3N_2$ index computed from these maps. The aperture used in the DAP is shown in orange on the gas-phase metallicity panel of Fig~\ref{fig:manga-maps}. We use this $\log Z = 8.65$ gas-phase metallicity value to compute the luminosity-to-molecular-gas-mass conversion factor because this elliptical aperture encompasses most of the CO(1-0) emission. A similar value of $\log Z = 8.67$ is given by the DAP from the same method with a $2.5''$ radius aperture, also shown in  Fig~\ref{fig:manga-maps}.

As the gas-phase metallicity is close to the solar value, following \citet{genzel15}, we use, as in \citet{mazzilli21}, a correction to the Galactic value which is the geometric mean of the corrections established by \citet{bolatto13} and by \citet{genzel12}:
\begin{equation}\label{eq:alpha_co}
    \alpha_{\rm CO} = \alpha_G \sqrt{0.67 \times {\rm exp}(0.36 \times 10^{\rm 8.67 - log\,Z}) \times 10^{\rm -1.27\times(log\,Z - 8.67)}}.
\end{equation}
The obtained value is $\alpha_{\rm CO} = 4.41$~M$_{\odot}$~${(\rm K \, km \, s^{-1} \, pc^{2}})^{-1}$. The resulting molecular-gas surface density map is shown on Fig.~\ref{fig:ks-plots}. The total observed mass of molecular gas is $M_{\rm mol \, gas \, obs} = 4.2 \, 10^{10}$~M$_{\odot}$\footnote{Using the gas-phase metallicity map, one may try to compute a spatial grid of conversion factors, assuming the dependency of the conversion factor on metallicity of \citet{genzel15}, even if the latter applies to whole galaxies. Such a map is shown on the last panel of Fig.~\ref{fig:manga-maps}. With this spatially-dependent conversion factor, a molecular-gas mass very close to the one assuming a uniform conversion factor is obtained: $M_{\rm mol \, gas } = 4.1 \, 10^{10}$~M$_{\odot}$  ($M_{\rm H_2 } = 3.1 \, 10^{10}$~M$_{\odot}$). } ($M_{\rm H_2 } =3.1 \, 10^{10}$~M$_{\odot}$, as in \citet{mazzilli21}). 

The conversion factor may differ significantly from the Galactic value if the gas is in a different physical state or has a large column density \citep{bolatto13}. For example, the typical value assumed in Ultra-Luminous Infrared Galaxies (ULIRGs) is 0.8~M$_{\odot}$~${(\rm K \, km \, s^{-1} \, pc^{2}})^{-1}$ \citep{1998ApJ...507..615D}, meaning that a lower mass of molecular gas is estimated from a given CO emission. Our system is likely composed of two galaxies with their molecular gas potentially in different states, and there may be outflows, making it all the more difficult to estimate a reliable conversion factor.

\subsection{Star formation rate from MaNGA H$\alpha$}

\begin{figure*}
\centering
    \includegraphics[width = 0.75 \linewidth]{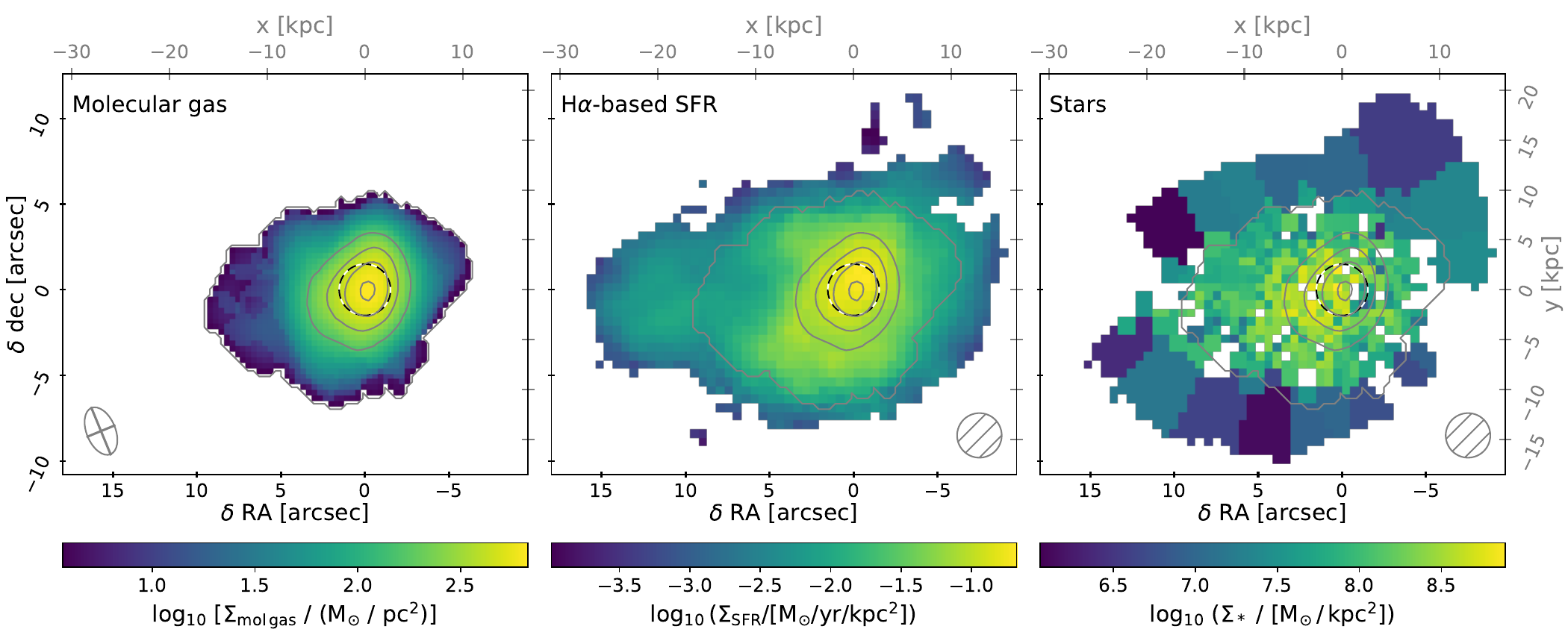} 

    \includegraphics[width = \linewidth]{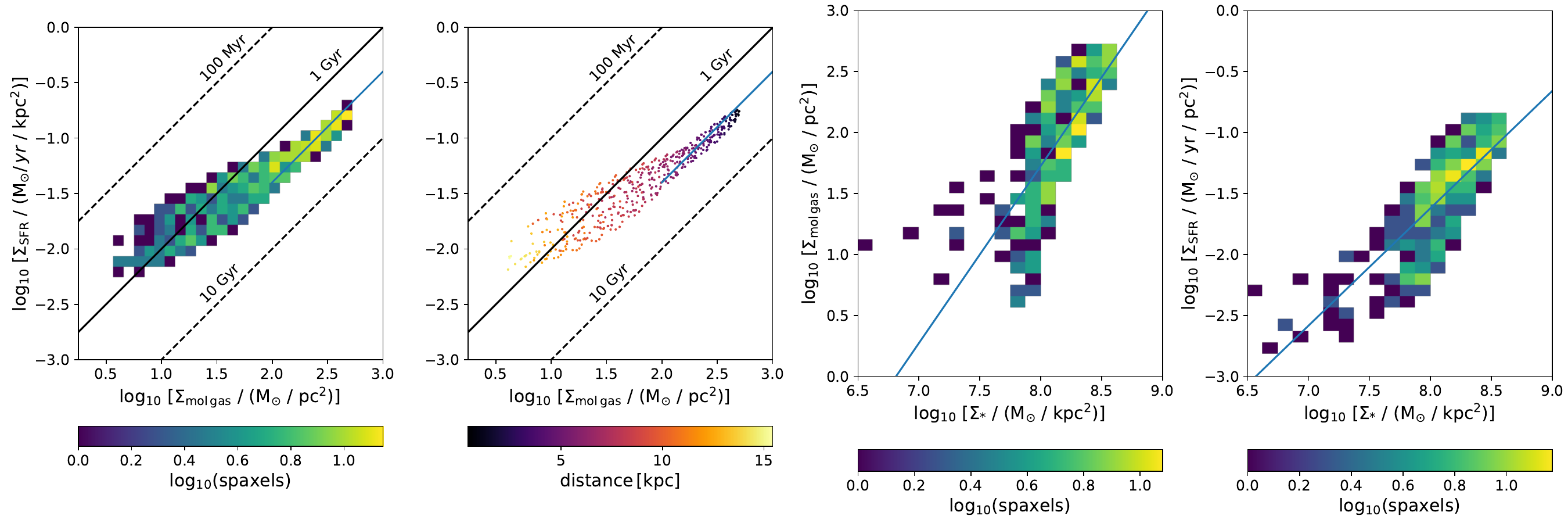} 
    \caption{Top row: Molecular gas surface density (left), H$\alpha$-based SFR surface density (middle) and stellar surface density (right). CO(1-0) intensity contours are over-plotted in grey. The 3$''$ SDSS fibre is over-plotted in a dashed black circle. Bottom row: diagrams obtained after degrading the maps to the same spatial resolution (see text). From left to right: KS diagram with 2D bins colour-coded by the number of contributing MaNGA spaxels, KS diagram with one dot per MaNGA spaxel, colour-coded by its distance to the centre of the SDSS fibre, molecular-gas surface density versus stellar surface density, and SFR surface density versus stellar surface density. On the two KS diagrams, black diagonal lines of constant depletion times are over-plotted as well as a blue line of depletion time the average for high-density molecular gas.}
    \label{fig:ks-plots}
\end{figure*}

The SFR is obtained from the H$\alpha$ luminosity similarly to the derivation in \citet{mazzilli21}. We compute an H$\alpha$ based SFR using \citet{1998ARA&A..36..189K}:
\begin{equation}
{\rm SFR(H\alpha)}  =  7.9 \, 10 ^{-42} \, \frac{L_{\rm corr}(\mathrm{H\alpha})}{\rm erg / s} \mathrm{M_{\odot}/yr},
\end{equation}
which assumes a Salpeter initial mass function. $L_{\rm corr}(\rm H\alpha)$ is the luminosity of the H$\alpha$ emission-line corrected from dust attenuation. $L_{\rm corr}(\rm H\alpha)$ is obtained by correcting the observed luminosity $L_{\rm obs}\rm (H\alpha)$ by:
\begin{equation}
    L_{\rm corr}(\mathrm{H\alpha}) =  L_{\rm obs}(\mathrm{H\alpha})\, 10^{0.4 \, A_{\lambda_{\rm H\alpha}}} = L_{\rm obs}(\mathrm{H\alpha})\, 10^{0.4 \, k(\lambda_{H\alpha}) \, E(B-V)},
    \label{eq-dustcorr}   
\end{equation}
where $A_{\lambda_{\rm H\alpha}}$ is the attenuation at the H$\alpha$ rest-frame wavelength, $k(\lambda_{\rm H\alpha})$ is the value of the total attenuation curve at the H${\alpha}$ rest-frame wavelength, and $E(B-V) = A_B - A_V$ is the colour excess, with $A_B$ and $A_V$ the attenuations in the $B$ and $V$ bands (respectively). The total attenuation curve is parametrised for example by \citet{Odonnel94} in the Milky Way, with a ratio of total to selective extinction in the $V$~band $R_V= \frac{A_V}{E(B-V)} = 3.1$. Using Eq.~\ref{eq-dustcorr} applied to both H$\alpha$ and H$\beta$, the colour excess is computed by:
\begin{equation}
    E(B-V) = \dfrac{2.5}{k(\lambda_{\rm H\beta}) - k(\lambda_{\rm H\alpha)})} \, \mathrm{log}_{10} \left[ \frac{\rm (H\alpha / H\beta)_{obs}}{2.86}\right],
    \label{eq-ebmv}
\end{equation}
where ${\rm (H\alpha / H\beta)_{obs}}$ is the Balmer decrement, the ratio of observed H$\alpha$ and H$\beta$ luminosities (both affected by dust absorption), and an intrinsic (with no effect from dust) Balmer decrement H$\alpha$/H$\beta$ of 2.86 is assumed, which corresponds to a temperature of ${\rm T=10^4\,K}$ and an electron density of $n_e=10^2{\rm cm^{-3}}$, for a case B recombination \citep{Osterbrock2006}. This computation of SFR by a correction of H$\alpha$ luminosity may not correct for the whole effect of dust absorption if Lyman continuum photons emitted by stars are directly absorbed by dust and therefore not accounted for.  

The map of the colour excess $E(B-V)$ is shown in Fig.~\ref{fig:manga-maps}. It is computed from the H$\alpha$ and H$\beta$ flux maps of the same figure, and from the attenuation curve of \citet{Odonnel94} with $R_V=3.1$. The map of SFR surface-density is shown in Fig.\ref{fig:ks-plots}. A total SFR of 19~M$_{\odot}$/yr is obtained\footnote{Assuming a Kroupa IMF \citep{kroupa01} instead of the Salpeter IMF would lower this value by a factor 1.6 (considering masses of stars range from 0.1 to 100~M$_{\odot}$, and assuming that only stars with masses larger than 20~M$_{\odot}$ contribute to the ionisation seeding the H$\alpha$ flux).}. Removing the AGN contribution $f_{\rm AGN}$ to the H$\alpha$ flux estimated by the method of \citet{Jin2021} gives an only slightly lower total SFR of 18~M$_{\odot}$/yr\footnote{The map of this contribution $f_{\rm AGN}$ is shown in Fig.~\ref{fig:manga-maps}. It is computed from the three line ratios shown in the same figure Fig.~\ref{fig:manga-maps}.}. 

\subsection{Resolved galaxy baryonic content and star-formation diagnostics combining NOEMA and MaNGA}

The angular resolutions of both MaNGA and NOEMA observations (spaxel size and PSF or beam FWHM) are given in Table~\ref{tab:obs-resolution}. To compare the surface densities of SFR and  molecular-gas, we convolve the molecular-gas  map by the NOEMA beam rotated by $90^{\circ}$, to emulate  a circular beam $B_{\rm common}$ of FWHM $\sqrt{l_{\rm major}^2 + l_{\rm minor}^2}$, with $l_{\rm major}$ (resp. $l_{\rm minor}$) the FWHM of the NOEMA major (resp. minor) axis. We also convolve the MaNGA maps by a circular Gaussian beam of FWHM $\sqrt{l_{\rm major}^2 + l_{\rm minor}^2 - l_{\rm PSF}^2}$, with $l_{\rm PSF}$ the FWHM of the MaNGA PSF, to also emulate a convolution with $B_{\rm common}$.
Spatial resolution is thus degraded in both maps to the same beam. We then reproject the molecular gas map on the MaNGA bins\footnote{Gaussian kernels are built for the NOEMA beam and the MaNGA PSF using the \texttt{radio-beam} package, convolution is performed with \texttt{astropy} \citep{astropy:2013}, and reprojection with the \texttt{reproject} package, using a bilinear interpolation.}. The resulting Kennicutt-Schmidt (KS) diagram, showing $\Sigma_{\rm SFR}$, as a function of $\Sigma_{\rm mol \, gas}$, is shown on Fig.~\ref{fig:ks-plots}. The left version of the diagram represents the number of MaNGA spaxels falling into a 2D bin of the KS space, and the right version represents dots corresponding to MaNGA spaxels, colour-coded by their distance to the centre of the SDSS fibre. Black lines show locations of constant molecular-gas depletion times $t_{\rm dep} = \dfrac{\Sigma_{\rm mol \, gas}}{\Sigma_{\rm SFR}}$ of 100~Myr, 1~Gyr and 10~Gyr. The average depletion time of the densest central parts, computed as the average of depletion times of MaNGA spaxels with a surface density larger than 100~M$_{\odot}$/pc$^2$, is of 2.5~Gyr, and the corresponding star-formation efficiency ${\rm SFE}= \dfrac{1}{t_{\rm dep}} = \dfrac{\Sigma_{\rm SFR}}{\Sigma_{\rm mol \, gas}}$ is $0.04/[100\,{\rm Myr}]$ (4$\%$). The larger scatter in the KS diagrams at low molecular-gas surface density may, beside the uncertainty in $\alpha_{\rm CO}$, reflect the more likely under-estimation of CO(1-0) flux and higher relative error in regions where the CO(1-0) flux densities are close to the spatially-uniform threshold value adopted. The surface-densities of molecular gas and SFR are also shown as a function of the stellar density obtained from the MaNGA value-added catalogue \texttt{Firefly} \citep{2022MNRAS.513.5988N, 2017MNRAS.466.4731G}. These surface-densities appear to be correlated (with slopes of 1.5 and 1 of fits, shown as blue lines, obtained by linear regressions for the molecular gas and SFR, respectively), similarly to the results on a molecular-gas main sequence and a resolved star-formation main-sequence obtained in \citet{lin19}. The correlation found for all the diagrams of Fig.~\ref{fig:ks-fit} however partly comes from the large size of $B_{\rm common}$ (FWHM of 3.4$''$) compared to the angular size of the CO(1-0) emission.

The surface densities used in the bottom row of Fig.~\ref{fig:ks-plots} are not corrected for inclination because of the complexity of the system, likely consisting of two superposed discs with different inclinations. The axis ratio of the Petrosian analysis of the $r$-band provided in the MaNGA DAP catalogue is $0.74$. The inclination deduced from this ratio is about 43$^{\circ}$ by modelling the system as an oblate spheroid with an intrinsic axis ratio $q_0$ between 0 and 0.2 (so that the inclination $i$ satisfies $\cos^2 i = \dfrac{\left(b/a\right)^2 - q_0^2}{1-q_0^2}$ \citep{hubble26}). Using this value would shift all plotted quantities by the same amount of -0.13~dex (in particular, this would not impact estimates of depletion times).

\section{Modelling the system with two disc galaxies}
\label{sec:model}

In the following we discuss the double-Gaussian fitting method of MaNGA emission-lines used to separate the system into two discs in \citet{mazzilli21}, and then present a modelling with two input discs whose parameters are constrained by a comparison to the observations. We start by applying this method to the molecular-gas, which is dynamically colder, and whose observations have a better spectral resolution, in order to constrain parameters. We then use the obtained parameters to separate the contributions of the two discs to the MaNGA spectra. SFE diagnostics and ionisation properties for the two individual modelled discs are then obtained.

\subsection{Single and double-Gaussian fits of MaNGA spectra}

\begin{figure}
    \centering
    \includegraphics[width=\linewidth]{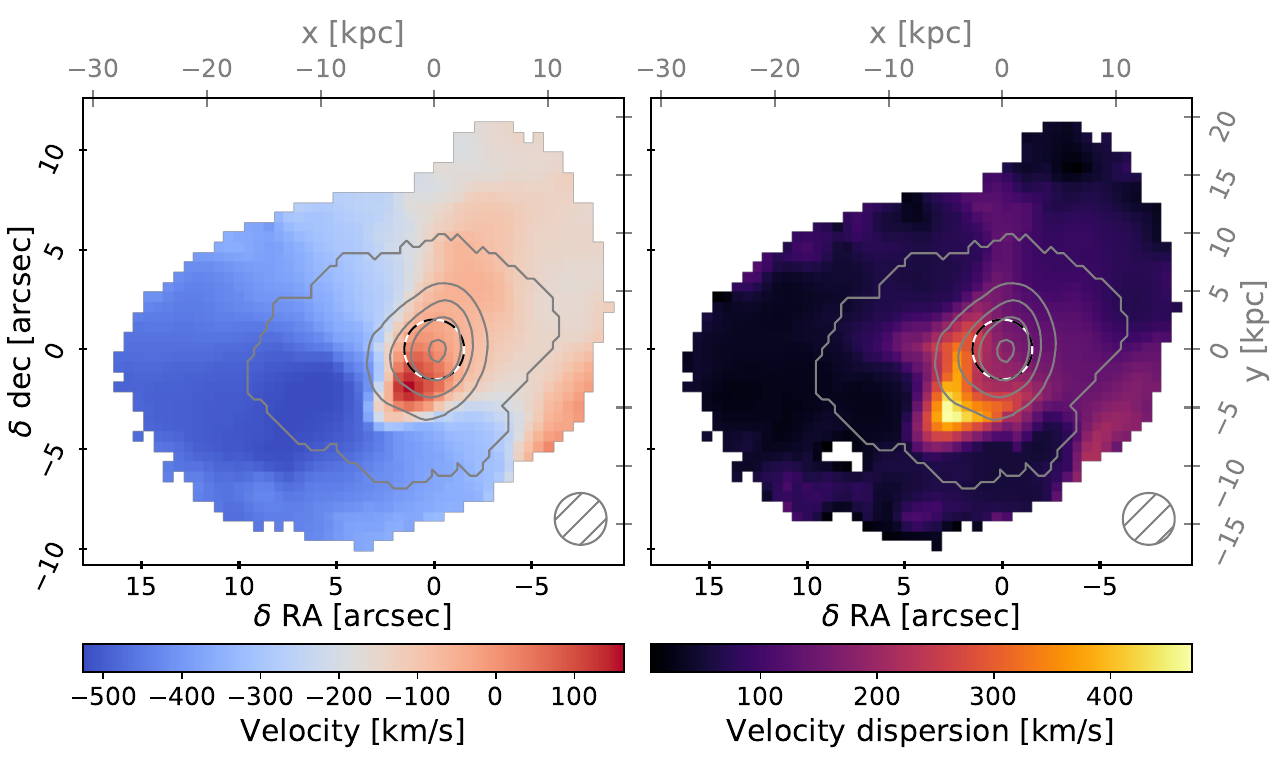}
    \caption{Velocity and velocity dispersion of H$\alpha$ from the Gaussian fit procedure.}
    \label{fig:iongas-vel}
\end{figure}

Figure~\ref{fig:iongas-vel} shows the velocity and velocity maps of ionised gas obtained with the procedure used in \citet{mazzilli21}, in its single-Gaussian version. At each spaxel, emission-lines (Balmer and other lines) are constrained to have the same central velocity, and the same velocity dispersion (removing the instrumental line-spread-function (LSF) effect), and this fit is performed on the whole galaxy with constraints on the variation of amplitude, central velocity and velocity dispersion coming from the values in neighbouring spaxels (as explained in Sec.\ref{sec:pres}, we use the now available MaNGA DR17 HYB10-MILESHC-MASTARSSP cube instead of the DR16 VOR10-GAU-MILESHC cube used in \citealt{mazzilli21}). This determination of velocity for ionised gas through Gaussian-fitting is different from the determination for molecular gas from the computation of the first moment, making a direct comparison between the ionised-gas velocity map and the molecular gas velocity map of the middle panel Fig.~\ref{fig:noema-moments} difficult. Both maps however show the same feature of the highest-velocity gas at and around the location of the SDSS fibre. Ionised gas velocity dispersion is larger than the molecular gas one as expected for a warmer component (as for the velocity, the determination is also different than for the molecular gas). 

In \citet{mazzilli21}, the decomposition into two discs through the double-Gaussian fitting of ionised-gas emission-lines procedure allowed to isolate emission-line fluxes maps and velocity maps for the two individual components. The assumption made for this decomposition was that for each spaxel where a double-Gaussian fit was better then a single-Gaussian fit (according to a $F$-test), each of the two Gaussians corresponded to one of the two galaxies. For spaxels in which a single-Gaussian fit was better, the Gaussian was assumed to correspond to the larger galaxy. A fitting procedure of the stellar component also allowed to separate two distinct stellar components. 
The inclinations of the gas discs were estimated from the shapes of ellipses fitting emission-line maps. The lengths of the minor and major axes were not very well defined, especially for the secondary galaxy whose maps had shapes departing significantly from an ellipse. The rotation velocities $V_{\rm rot}$ were estimated by $V_{\rm rot} = \dfrac{W}{2 \sin(i)}$, with $W$ defined as the difference between the 90$^{\rm th}$ and 10$^{\rm th}$ percentile points of the velocity histogram, and $i$ the previously determined inclination. The dynamical masses were estimated by $M_{\rm dyn} = V_{\rm rot} R / G $ with $R$ the maximum radius where H$\alpha$ is detected. The parameters found by \citet{mazzilli21} are listed in Table~\ref{tab:params-gals-m21}.

\begin{table}[]
    \caption{Parameters found by \citet{mazzilli21}}
    \centering
    \begin{tabular}{l|c|c}
    \hline 
    \hline 
    galaxy & main & secondary \\
    \hline 
        inclination & $[49^{\circ}, 53^{\circ}]$ & $[46^{\circ}, 50^{\circ}]$  \\
        stellar mass [M$_{\odot}$] & $1.43 \, 10^{11}$ & $1.59 \, 10^{10}$  \\
        rotation velocity [km/s] & $236$ & $140$ \\
        dynamical mass [M$_{\odot}$] & $2.65 \, 10^{11}$ & $2.93 \, 10^{10}$ \\
    \hline  
    \end{tabular}
   
    \label{tab:params-gals-m21}
\end{table}

The double-Gaussian fitting-procedure is not always able to properly separate the two discs. Double-peaks can appear in spectra of spaxels near the centre of discs because of rotation \citep[e.g.][]{maschmann23}. The large velocity dispersion of ionised gas and the widening of emission-lines by the MaNGA LSF worsens the problem. 

\subsection{Modelling from the CO(1-0) NOEMA observations}

\begin{figure}
    \centering
    \includegraphics[width = \linewidth]{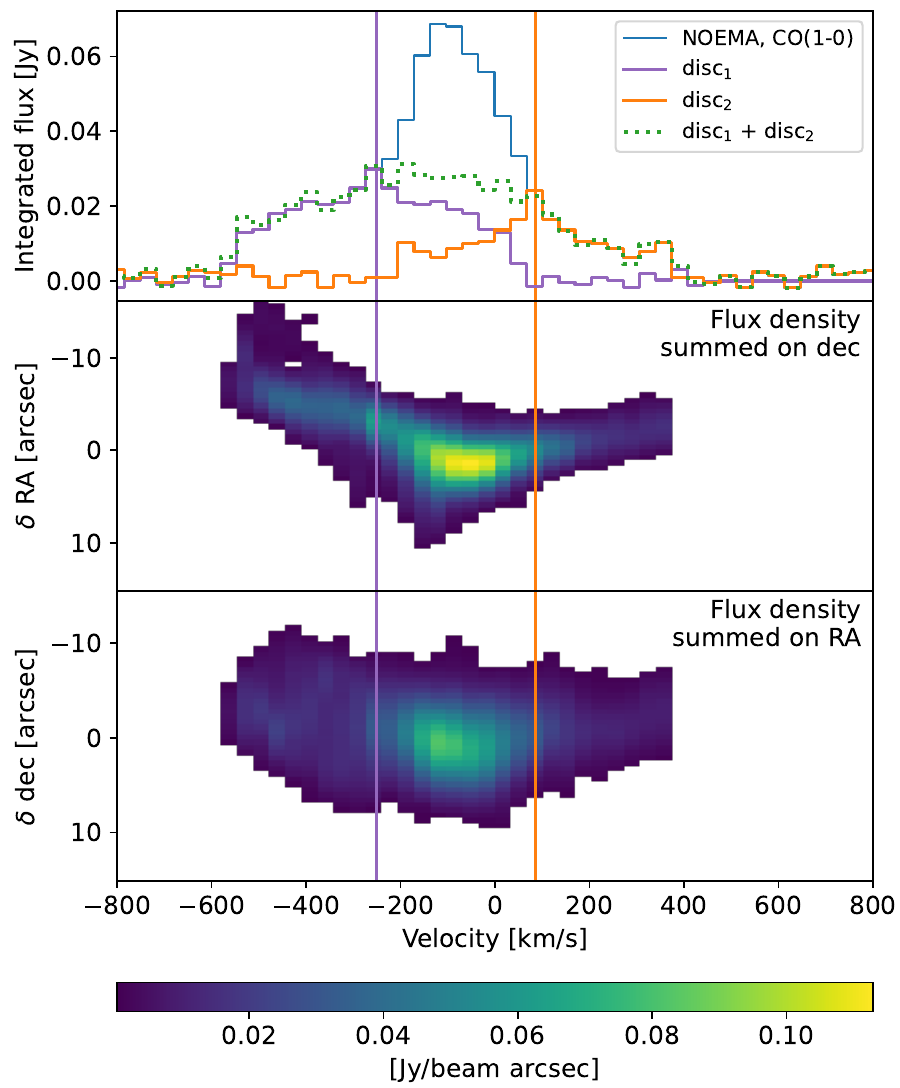}
    \caption{Attempt of reconstructing the observed spectrum, assuming there are two discs, that each one has a half part accounting for the CO(1-0) emission at the lowest or highest observed velocities, and that the CO(1-0) emission of each disc is symmetric w.r.t. its minor kinematic axis. The violet (resp. orange) line is an example of systemic velocity for the first (resp. second) disc, with the spectrum of the disc in the same colour in the top panel. Top: NOEMA spectrum, discs spectra reconstructed by symmetry and their sum. Middle: NOEMA cube integrated on declination. Bottom: NOEMA cube integrated on RA.}
    \label{fig:recons-sym}
\end{figure}

\begin{figure}
    \centering
    \includegraphics[width = \linewidth]{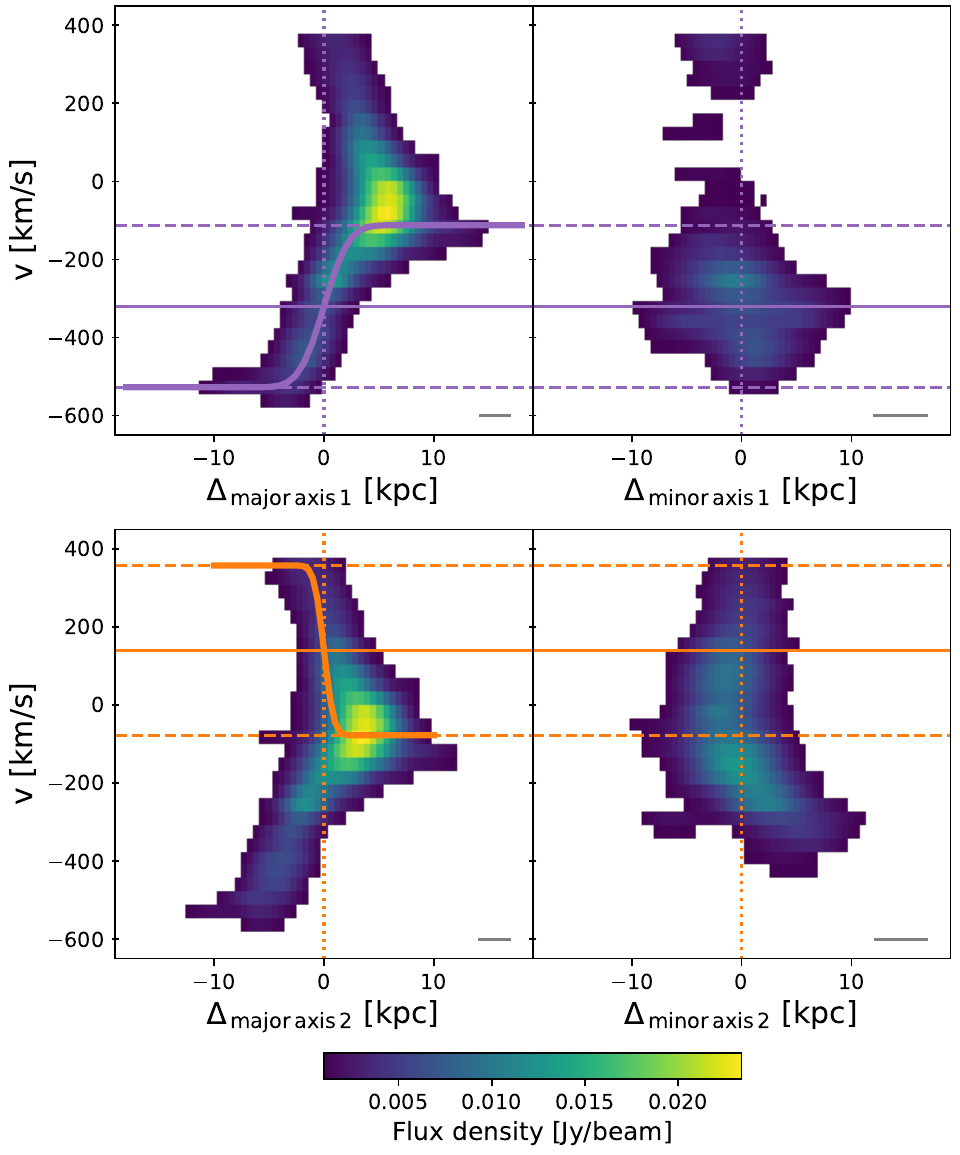}
    \caption{Position-velocity diagrams of NOEMA observations, centred on the estimated kinematic centre of disc$_1$ (first row) or on the estimated kinematic centre of disc$_2$ (second row) and computed on a one-spatial-pixel-wide slice either along the major axes of discs (first column) or the minor axes (second column). $v_c(R) \sin(i)$ curves are represented on diagrams with position taken along the major axis (first column). Horizontal solid lines show the fitted systemic velocities $v_{\rm sys}$ of the discs and horizontal dashed lines show $v_{\rm sys} \pm V_{\rm max} \sin (i)$. Violet lines correspond to disc$_1$, and orange curves to disc$_2$. The horizontal grey lines at the bottom-right corner of each panel is the projection of the elliptical Gaussian NOEMA synthesised beam on the $x$-axis of the panel.}
    \label{fig:pv-co-fit}
\end{figure}

\begin{figure}
    \centering
    \includegraphics[width = \linewidth]{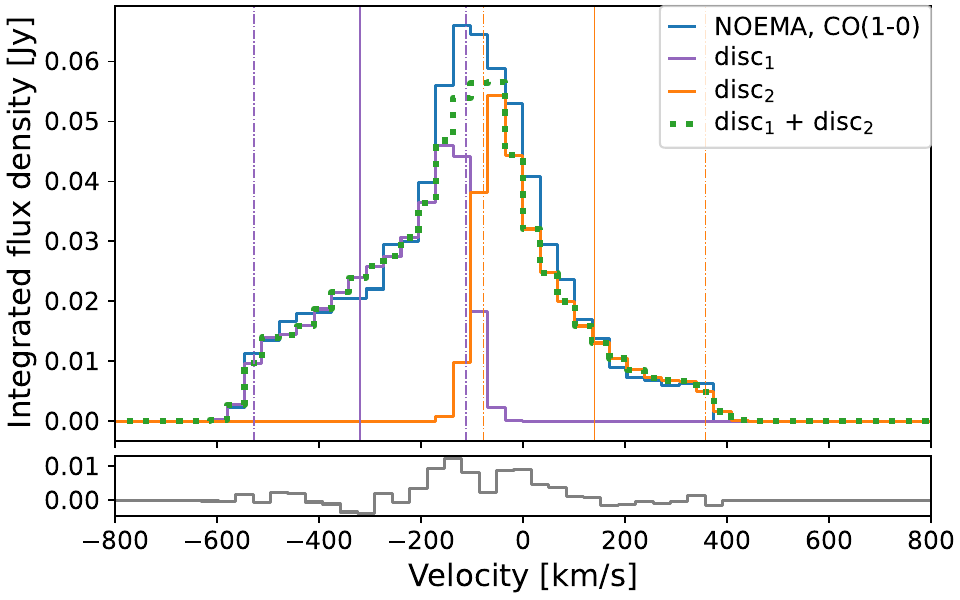}
    \caption{Observation and model of CO(1-0) emission integrated on the system. Vertical solid lines show the fitted systemic velocities $v_{\rm sys}$ of the discs and vertical dashed lines show $v_{\rm sys} \pm V_{\rm max} \sin (i)$.}
    \label{fig:total-co-fit}
\end{figure}

\begin{figure*}
    \centering
    \includegraphics[width = \linewidth]{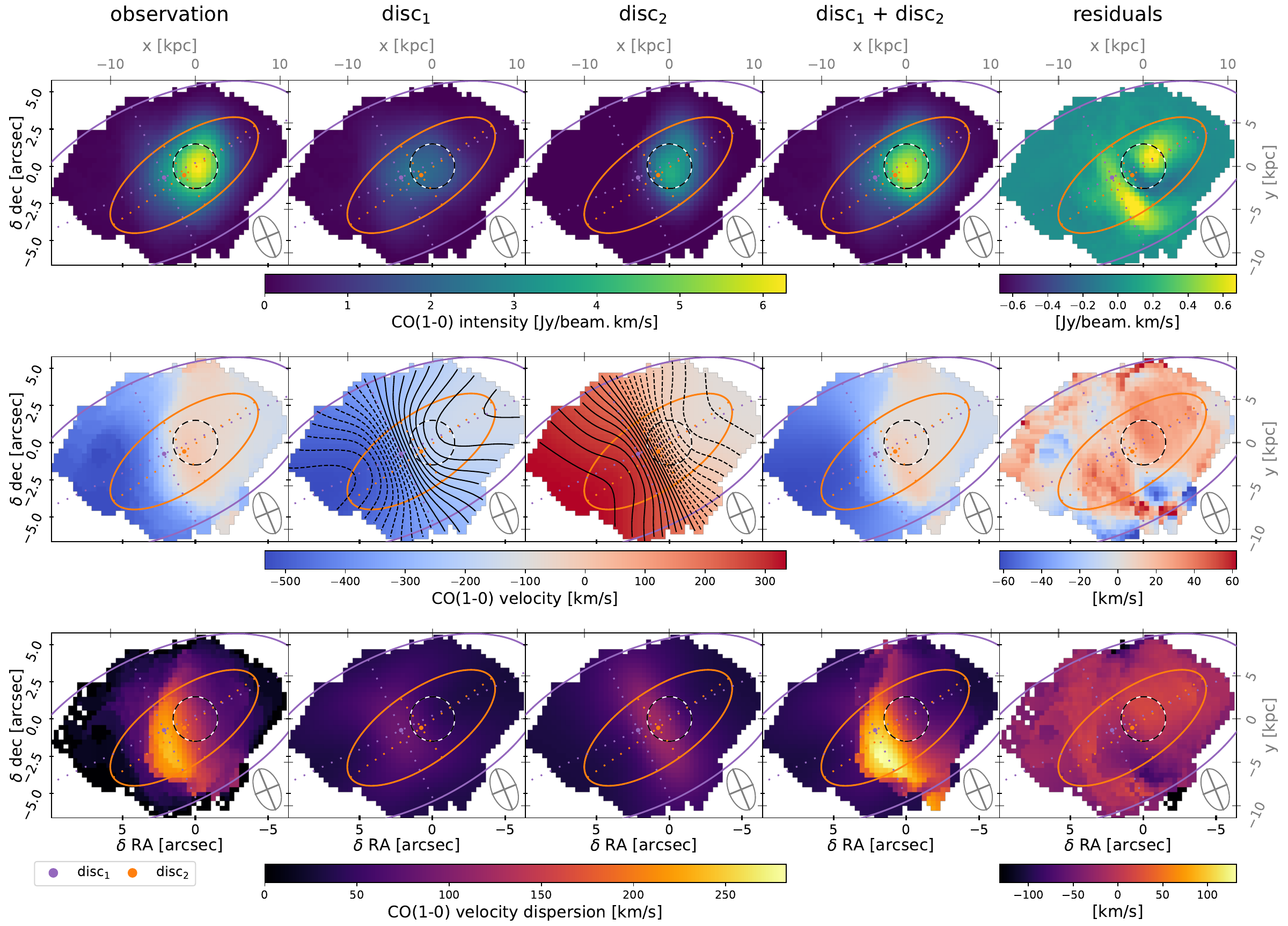}
    \caption{Observations, models and residuals of intensity, velocity and velocity dispersion from the CO(1-0) line. Models are shown both for each disc (column 2 and 3) and for the two superposed discs (column 4). Residuals are the observation (column 1) minus the two superposed discs (column 4). On the velocity maps of individual discs, iso-line-of-sight velocity curves are represented in black dashed (resp. solid) lines for values below (resp. above) the systemic velocity of the disc.}
    \label{fig:moms-res-co-fit}
\end{figure*}

From a simple analysis illustrated in Fig.~\ref{fig:recons-sym}, it appears that if the CO(1-0) emission originates from two superimposed rotating discs, the molecular gas cannot be axisymmetrically distributed in each disc with an axisymmetric CO-to-molecular gas conversion factor in each disc. In the case of an axisymmetric distribution of molecular gas and an axisymmetric conversion factor, the spectrum is symmetric around the velocity of the centre of mass of the galaxy (systemic velocity). The position-velocity diagrams (shown in Fig.~\ref{fig:recons-sym}, integrated on declination or RA), in which the characteristic shapes of emission of discs in rotation can be identified, allow us to isolate two parts of the velocity distribution which are most likely due to half of only one of the supposed discs: the lowest and highest velocity parts of the spectrum. By reconstructing the spectrum of each galaxy by symmetry w.r.t. an assumed systemic velocity, and by then summing the two obtained spectra, it is not possible to obtain the observed spectrum for any values of systemic velocities such that the extent of the individual spectra imply reasonable ($<300$~km/s) rotation velocities. There is too much emission from gas at intermediate velocities. Some clumpy distribution of the molecular gas may explain this asymmetry. Some variation in metallicity or in the physical properties of the molecular gas may also make the conversion factor vary. Both molecular discs or one of them may be disturbed by a tidal interaction, resulting in an asymmetric distribution of the molecular gas. Some tidal features such as a tail or a bridge may be present. Finally, there may be some molecular outflow.

\begin{table}[]
    \caption{Parameters for models}
    \centering
    \begin{tabular}{l|c|c}
    \hline
    \hline
         Galaxy & disc$_1$ & disc$_2$ \\
         \hline
        $v_{\rm sys, \, disc}$ [km$/$s] & -320 & 140 \\
        inclination & $60^{\circ}$ & $65^{\circ}$ \\
        PA & $295^{\circ}$ & $125^{\circ}$ \\
        $R_0$ [kpc] & 2.5 &  1 \\
        $a$ [kpc] & 4.5 & 2.5 \\
        $V_{\rm max}$ [km/s] & 240 & 240 \\
\hline
    \end{tabular}
    \label{tab:params-gals}
\end{table}

We realise mock NOEMA cubes with two molecular-gas discs superimposed on the line-of-sight. We first generate discs from particles of positions and velocities adjusted to a set of chosen parameters. We model gas discs with a Miyamoto-Nagai density profile:
\begin{equation}\label{eq:mn}
\rho_d(R,z) = \left( \dfrac{h^2 M}{4 \pi} \right) \dfrac{a R^2 + (a + 3\sqrt{z^2 + h^2}) (a + \sqrt{z^2 + h^2})^2}
    {\left[ R^2 + (a + \sqrt{z^2 + h^2})^2\right]^{\frac{5}{2}} (z^2 + h^2)^{\frac{3}{2}}},
\end{equation}
where $M$ is the total mass of the disc, $a$ is a radial scale length, and $h$ is a vertical scale length. We set a vertical scale length $h=100$~pc for each disc. We model each rotation curve with a close to linear rise from $R=0$ and a smooth transition to a plateau, using an error function: the circular velocity is $v_c(R) = V_{\rm max} \mathrm{erf} (R/R_0)$, where $V_{\rm max}$ is the plateau velocity and $R_0$ is the transition radius from the initial rise to the plateau. The two parameters describing each rotation curve are thus the transition galactocentric radius from rise to plateau, $R_0$, and the velocity of the plateau, $V_{\rm max}$. The rotation curves are shown in Fig.~\ref{fig:pv-co-fit}. 
For each disc, we set the molecular-gas radial and azimuthal velocity dispersions ($\sigma_R$ and $\sigma_{\phi}$, respectively) to the same value $\sigma_R = \sigma_{\phi} = 30$~km$/$s, independent of radius for simplicity, and set the vertical velocity dispersion $\sigma_z$ to $\sigma_z = 15$~km$/$s. We then bin the on-sky positions and line-of-sight velocities with the same spatial and velocity binning as for the observed NOEMA cube, and we then convolve each channel map with a two-dimensional elliptic Gaussian kernel with the FWHMs of the major and minor axes of the NOEMA synthesised beam and its position-angle, using \texttt{radio-beam} and \texttt{astropy} for this convolution. The value of $M$ in Eq.~\ref{eq:mn} is set to half the total observed molecular-gas mass $M_{\rm mol \, gas \, obs }$ determined in Sec.~\ref{sec-noemaobs} and the simulated molecular-gas mass per cube cell is converted into CO(1-0) flux density using the same conversion factor as the one used to compute $M_{\rm mol \, gas \, obs }$. The simulated flux densities are thus of the same order of magnitude as the observed ones. The simulated spectra of each disc are then adjusted spaxel by spaxel as described below.

We consider a varying contribution of disc$_1$ and disc$_2$, the ``main galaxy'' and ``secondary galaxy'' of \citet{mazzilli21} (respectively), in each spaxel. With $s_1$ and $s_2$ the mock spectra of disc$_1$ and disc$_2$, respectively, in a given spaxel, we fit two multiplying factors $f_1$ and $f_2$, one for each disc (disc$_1$ and disc$_2$, respectively) so that the sum of the two resulting spectra, each multiplied by the corresponding fitted factor, fits best the observed spectrum, i.e. so that $f_1 s_1 + f_2 s_2$ fits the observed spectrum. This resembles the approach used in \citet{2019A&A...623A..79C}, where a multiplicative filter is applied to the simulated cubes so that the spectrum in a spaxel is normalised to the observed spectrum in this spaxel. The fitting is performed after convolution of the mock cube by the NOEMA synthesised beam and the fitted factors must not vary strongly on the area of the synthesised beam for this approach to be physically valid (otherwise, the contribution of the molecular gas of a given disc at a given location would be inconsistent from one spaxel to its neighbours on which it is smeared by the spatial convolution). Because of the large number of parameters and the complexity induced by the superposition on the line-of-sight, we fit most parameters by eye, resorting to least-squares only for the local intensity of CO(1-0) emission per disc, i.e. only for the determination of $f_1$ and $f_2$ in each spaxel. Some of the parameters are listed in Table~\ref{tab:params-gals}. 

The NOEMA spectra are fitted after the same threshold cut as used to produce the moment maps. The NOEMA and fitted discs spectra stacked over the entire system are represented in Fig.~\ref{fig:total-co-fit}. Figure~\ref{fig:spectra-co-fit} displays the CO(1-0) spectra and models for spaxels stacked 16 by 16 all over the system. We represent maps of the intensity, velocity and velocity dispersion of the observations and of the model, with residuals, in Fig.~\ref{fig:moms-res-co-fit}. The first column shows the same maps as in Fig.~\ref{fig:noema-moments} except for the colour scale range, which is, for each row of Fig.~\ref{fig:moms-res-co-fit}, common to the fourth first maps. The second and third columns show the intensity, velocity and velocity dispersion of disc$_1$ and disc$_2$ (respectively), the fourth column shows the same quantities for the two modelled discs together, and the fifth column shows the residuals, computed as the difference between the first column (observations) and the fourth column (model). The colour scales of the residuals are centred on zero.

Molecular-gas masses can be computed, assuming for example the same $\alpha_{\rm CO}$ of Sec.~\ref{sec:mol-ion}. We find molecular-gas masses $M_{\rm mol \, gas \, disc_1 } = 2.1 \, 10^{10}$~M$_{\odot}$ for disc$_1$, and $M_{\rm mol \, gas \, disc_2 } = 1.8 \, 10^{10}$~M$_{\odot}$ for disc$_2$. The sum of these two masses is $M_{\rm mol \, gas \, disc_1 \, + \, disc_2 } = 3.9 \, 10^{10}$~M$_{\odot}$, slightly below the observed total mass $M_{\rm mol \, gas \, obs } = 4.2 \, 10^{10}$~M$_{\odot}$.  

\subsection{Modelling applied to the MaNGA observations}

\begin{figure*}
    \centering
    \includegraphics[width = \linewidth]{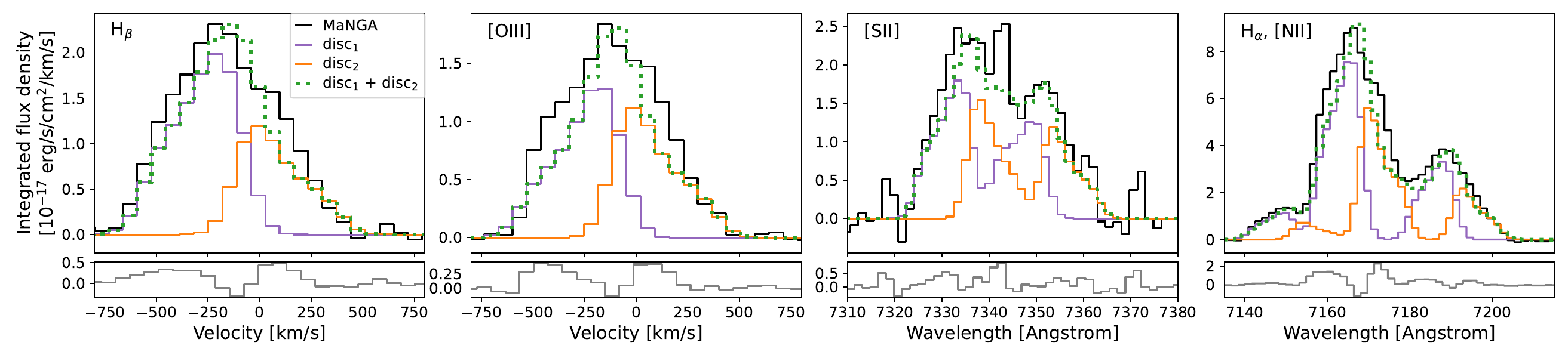}
    \caption{Observations and models of MaNGA emission-lines integrated on the system.}
    \label{fig:emlines-fit}
\end{figure*}

We realise mock cubes for a given emission-line. The velocity channels are convolved by a circular Gaussian kernel of FWHM 2.6 arcsec, accounting for the MaNGA PSF. We also perform a convolution on velocities with a 1D Gaussian kernel with the dispersion of the MaNGA line-spread-function at the redshifted wavelength of the considered emission-line and at each considered location (this dispersion is given in the MaNGA model cube). The conversion between wavelength $\lambda$ and velocities $v$ is done by:
\begin{equation}
    v = \dfrac{c}{1+z_{\rm source}} \left( \dfrac{\lambda - \lambda_0}{\lambda_0} - z_{\rm source} \right),
\end{equation}
where $\lambda_0$ is the rest-frame wavelength of the considered emission-line, and $z_{\rm source}$ is the redshift taken from the NASA Sloan Atlas catalogue. This allows us to compare modelled and observed spectra. We keep the same parameters as for the molecular-gas, except for velocity dispersion: we set the radial and azimuthal velocity dispersions to a larger value $\sigma_R = \sigma_{\phi} = 40$~km$/$s, and the vertical velocity dispersion to $\sigma_z = 20$~km$/$s. As for the molecular gas, we use a least-squares method to fit the contribution of each of the two discs to the spectra. The results, integrated on the system, are shown for a few emission-lines in Fig.~\ref{fig:emlines-fit}. In App.~\ref{app:oiii-two-discs}, we show the example of the [OIII]5008 fitting: a comparison of the observed and fitted moments in Fig.~\ref{fig:moms-res-oiii-fit}, and fits of spectra stacked 16 by 16 in Fig.~\ref{fig:spectra-oiii-fit}.

We compute the SFR of modelled disc$_1$ and disc$_2$ using the same dust correction as in Sec.\ref{sec:mol-ion}, and find a SFR of 11~$\mathrm{M_{\odot}/yr}$ for disc$_1$ and a SFR of 8~$\mathrm{M_{\odot}/yr}$ for disc$_2$, hence a total of 19~$\mathrm{M_{\odot}/yr}$, as the total observed SFR.

\subsection{Star formation and gas ionisation in the discs}

\begin{figure}
    \centering
    \includegraphics[width = \linewidth]{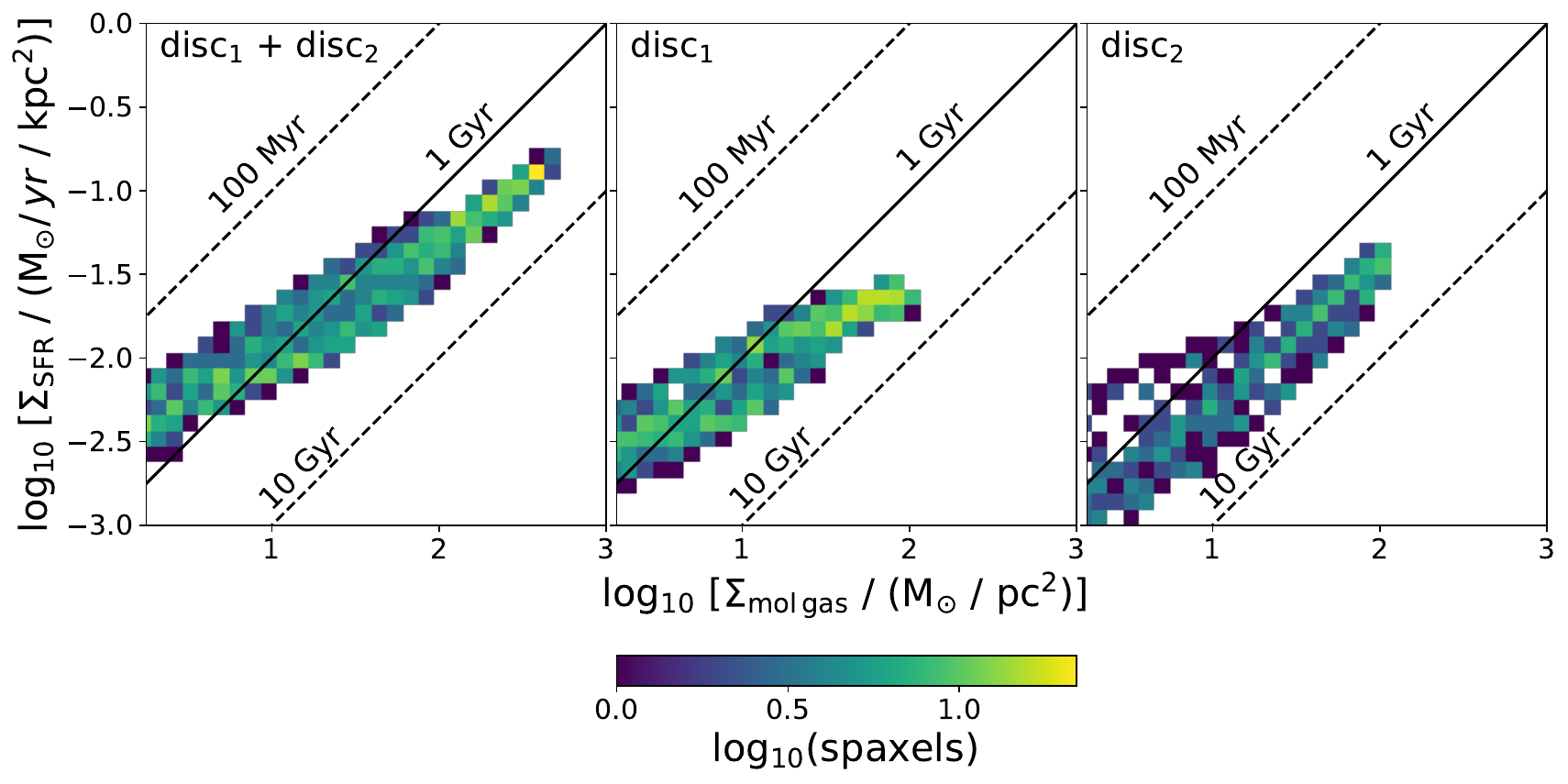}
    \caption{Kennicutt-Schmidt diagrams. Left: For the two superposed model discs, with no correction for inclination. Middle and right: For disc$_1$ (resp. disc$_2$), correcting for the modelled inclinations of 60 and 65 degrees, respectively.}
    \label{fig:ks-fit}
\end{figure}

Figure~\ref{fig:ks-fit} shows the Kennicutt-Schmidt diagrams of the two discs together, and of each individual disc (with a correction for inclination). The SFEs are similar but a little higher in the densest parts of disc$_1$ than in the densest parts of disc$_2$.

\begin{figure}

    \centering
    \includegraphics[width = \linewidth]{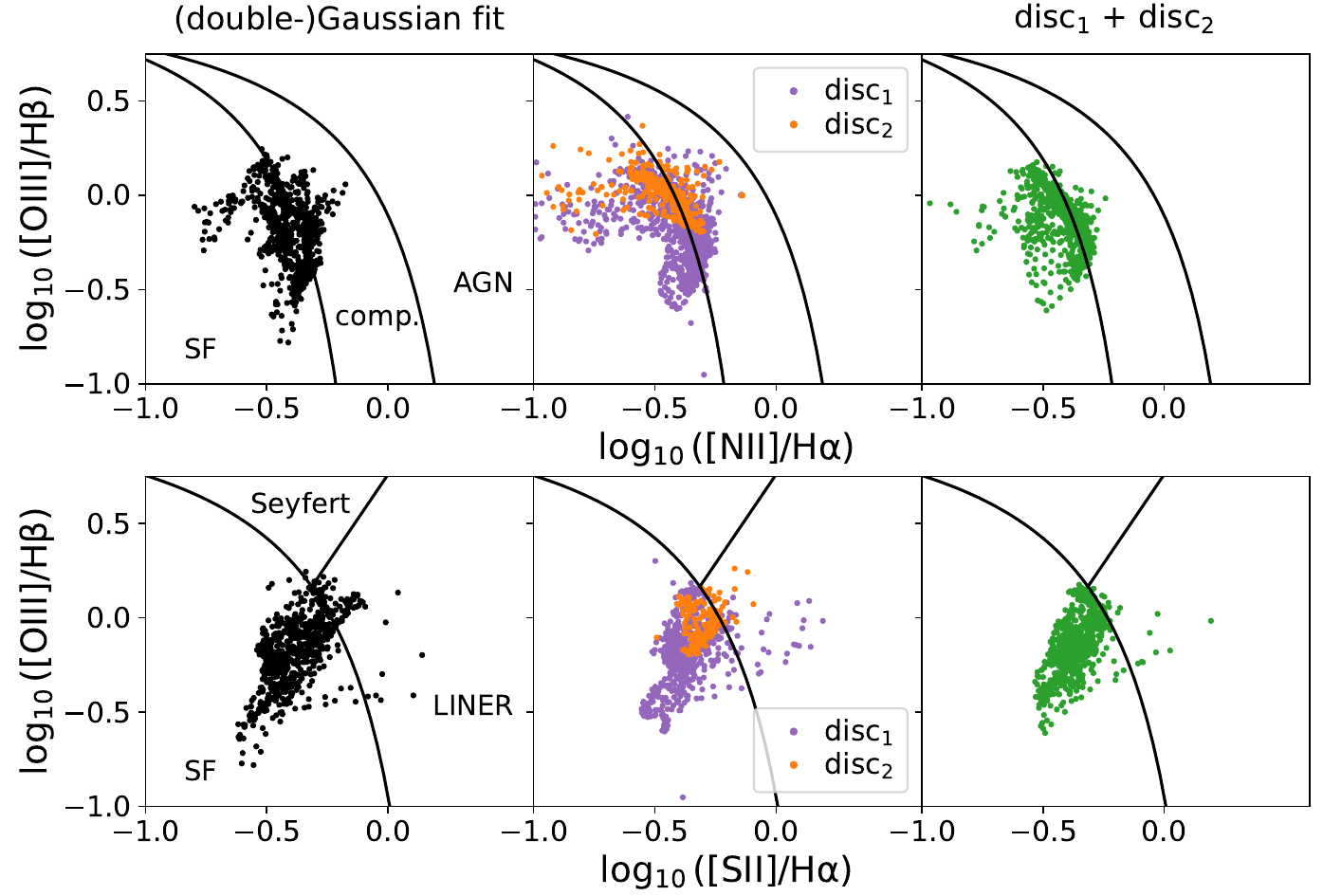}
    \caption{BPT diagrams. Each dot corresponds to a MaNGA spaxel. The zones separated by the lines are named on the left-column panels, with SF standing for star-forming, and ``comp.'' for composite.}
    \label{fig:bpt}
\end{figure}

Baldwin, Phillips and Terlevich (BPT) \citep{bpt81} diagrams are shown in Fig.~\ref{fig:bpt}, with curves defined by \citet{kauffmann03} and \citet{kewley01} separating different zones. Diagrams are shown for the double-Gaussian fit to the emission-lines which best determines the observed flux, for each modelled disc component, and for the sum of the two components. Each dot corresponds to a MaNGA spaxel. The ionisation in most of the spaxels seems to come from star-formation or from a composite origin (some AGN activity and/or shocks may contribute). The two discs mostly have a similar origin for their ionisation (their spaxels are in the same zones of the diagrams) but disc$_2$ has (as found by \citet{mazzilli21}), larger [OIII]/H$\beta$ ratios.

\section{Discussion}
\label{sec:discu}

Thanks to the high velocity resolution of NOEMA observations and the low velocity dispersion of molecular gas, we have been able to model two discs to the system J221024.49+114247.0. The main difference between our separation into two components and the one performed by \citet{mazzilli21} is the similar rotation velocities of the two discs. While the problem is degenerate and the exact values found in this work should be taken with caution, we conclude that the  rotation velocities of the discs indicate that the system is a major merger, rather than a minor merger, as was found previously. This holds at least for a modelling with two axisymmetric discs only. A determination of the dynamical mass of each galaxy can only be very approximate because the radial extent of the CO(1-0) emission is limited (compared to that of atomic hydrogen HI, ideal for rotation curves), and the shape of the ``secondary'' galaxy (disc$_2$) is uncertain. 

Decomposition of the MaNGA emission-lines is more difficult for several reasons: the intrinsic gas velocity dispersion is larger than for molecular gas, the instrumental LSF is large (with a dispersion $\simeq 70$~km/s), and the removal of the stellar contribution to the spectra adds uncertainty. A fitting of a model of two discs for both the stellar contribution and the gas contribution would be even more degenerate. It is thus much more difficult to fit any disc model to the MaNGA data than to the NOEMA data. The lower rotation velocity found by \citet{mazzilli21} for the ``secondary'' galaxy (disc$_2$) mainly comes from their assumption that the emission of spaxels best fitted by single-Gaussians was from the main galaxy, likely removing some of the low-velocity region of the secondary galaxy, and also from an SNR cut on the flux of the secondary galaxy, removing some of its high-velocity region. Resorting to a statistics on the velocity distribution of the spaxels cut at  both low and high velocities provided a lower rotation velocity than with our modelling by a disc.

The modelling presented here, while theoretically motivated (reproducing well-defined models of rotating discs), is conceptually too simple to capture the kinematics of the system. It considers two discs with neither non-axisymmetries, nor warps, nor tidal features and hence is bound to be imperfect, given the complex and perturbed optical images. The problem, because of the spatial and spectral overlap of galaxies, is already degenerate, and the size of disc$_2$ is only a few times the size of the NOEMA synthesised beam. We thus did not aim at reproducing the whole complexity of the system by adding even more parameters (describing warps, for example), but at exploring how far we could go in the understanding of the system with this two simple discs modelling.

Further uncertainty in the study of NOEMA observations comes from the unknown value of the conversion factor $\alpha_{\rm CO}$, which may differ on average from one disc to the other, and vary across each disc. A number of works show that this factor is lower in mergers (see \citealt{bolatto13} and references therein), because of the large column densities, temperatures, and large velocity dispersions. The existence of an outflow is also possible, with, again, a different conversion factor to consider. Out-flowing gas, with a lower conversion factor than gas in discs, may contribute to the central part of the total spectrum. 

\section{Conclusion}
 
We have further investigated J221024.49+114247.0, a $z=0.09228$ system previously studied in \citet{mazzilli21} and \citet{mazzilli24}, through interferometric NOEMA CO(1-0) observations. The high velocity resolution of the NOEMA observations and the low velocity dispersion of molecular gas allowed us to constrain a model of two counter-rotating discs. We aimed to make the best of this simple two-discs model without any non-axisymmetric feature in discs, both for molecular and ionised gas, while the reality is very likely more complex. 

This revisited the system as being a major merger. We applied the model to the ionised-gas MaNGA emission-lines and were able to draw resolved Kennicutt-Schmidt diagrams for both discs. The CO-to-molecular-gas conversion factor however remains uncertain, especially for a system of interacting galaxies. 

While this system is complicated to dissect because of the degeneracy induced by the superposition on the line-of-sight, further observations may allow one to deepen our understanding of its physics: some resolved radio-continuum or infrared JWST observations would allow to further study its star formation and a potential obscured AGN. Finally, such a source will be detectable in HI with SKA.

\begin{acknowledgements}
This work is based on observations carried out under project number S21BQ001 with the IRAM NOEMA Interferometer [30~m telescope]. IRAM is supported by INSU/CNRS (France), MPG (Germany) and IGN (Spain). We thank all the IRAM people who contributed to the observations and helped with their reduction. Funding for the Sloan Digital Sky Survey V has been provided by the Alfred P. Sloan Foundation, the Heising-Simons Foundation, the National Science Foundation, and the Participating Institutions. SDSS acknowledges support and resources from the Center for High-Performance Computing at the University of Utah. SDSS telescopes are located at Apache Point Observatory, funded by the Astrophysical Research Consortium and operated by New Mexico State University, and at Las Campanas Observatory, operated by the Carnegie Institution for Science. The SDSS web site is \url{www.sdss.org}. SDSS is managed by the Astrophysical Research Consortium for the Participating Institutions of the SDSS Collaboration, including the Carnegie Institution for Science, Chilean National Time Allocation Committee (CNTAC) ratified researchers, Caltech, the Gotham Participation Group, Harvard University, Heidelberg University, The Flatiron Institute, The Johns Hopkins University, L'Ecole polytechnique f\'{e}d\'{e}rale de Lausanne (EPFL), Leibniz-Institut f\"{u}r Astrophysik Potsdam (AIP), Max-Planck-Institut f\"{u}r Astronomie (MPIA Heidelberg), Max-Planck-Institut f\"{u}r Extraterrestrische Physik (MPE), Nanjing University, National Astronomical Observatories of China (NAOC), New Mexico State University, The Ohio State University, Pennsylvania State University, Smithsonian Astrophysical Observatory, Space Telescope Science Institute (STScI), the Stellar Astrophysics Participation Group, Universidad Nacional Aut\'{o}noma de M\'{e}xico, University of Arizona, University of Colorado Boulder, University of Illinois at Urbana-Champaign, University of Toronto, University of Utah, University of Virginia, Yale University, and Yunnan University.
\end{acknowledgements}

\bibliographystyle{aa}
\bibliography{biblio}

\appendix

\section{Visualisation of the NOEMA cube}

\begin{figure}[h]
    \centering
    \includegraphics[width = 0.8 \linewidth]{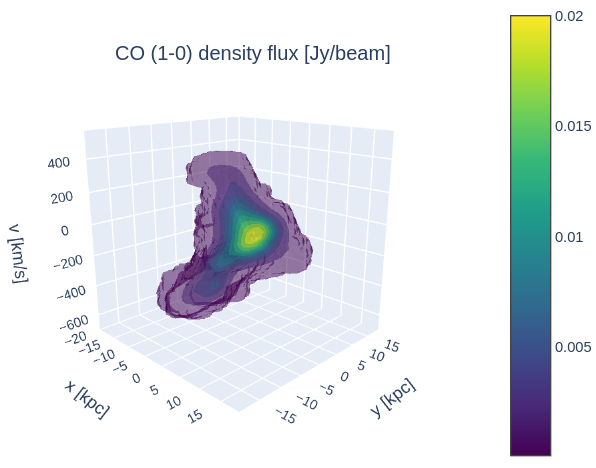}
    \caption{Visualisation of the NOEMA cube. This is a screenshot of the 3D widget of the cube made with plotly and available at 
\href{https://vm-weblerma.obspm.fr/~ahalle/noema_cube.html}{https://vm-weblerma.obspm.fr/$\sim$ahalle/noema$\_$cube.html}}
    \label{fig:noema-cube}
\end{figure}

\onecolumn
\section{Maps from MaNGA}
\label{app:manga}

\begin{figure*}[h!]
    \centering
    
    \includegraphics[width = \linewidth]{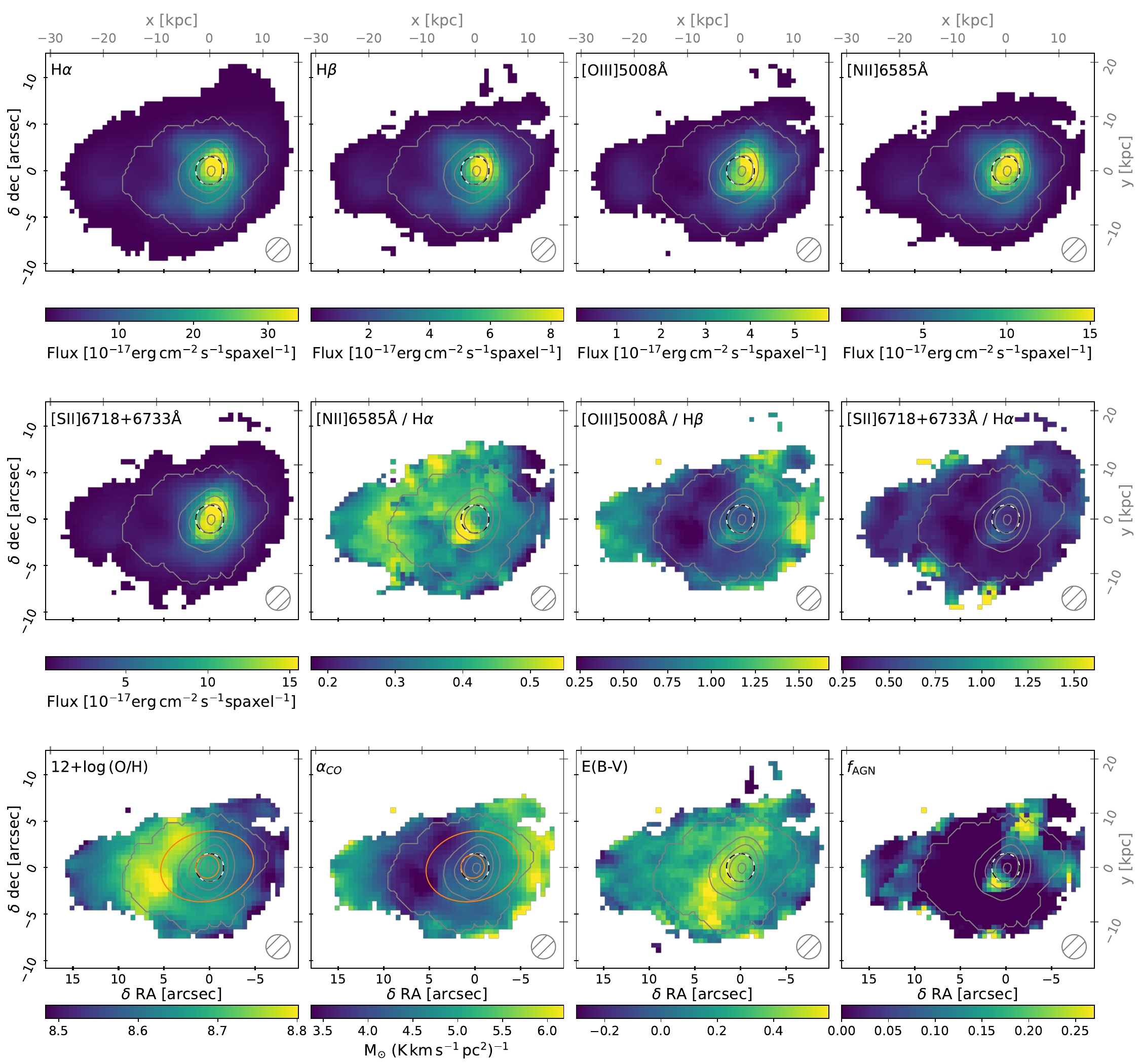}

    \caption{Maps obtained from the MaNGA data fitted by a double-Gaussian procedure. NOEMA CO(1-0) intensity contours are over-plotted in grey. The 3$''$ SDSS fibre is over-plotted in a dashed black circle. The elliptical aperture used to compute gas-phase metallicity is over-plotted in orange on the gas-phase metallicity map and the conversion factor map (see text).}
    \label{fig:manga-maps}
\end{figure*}

\onecolumn
\section{Two-disc modelling for CO(1-0)}
\label{app:co-two-discs}

\begin{figure*}[h!]
    \centering
    \includegraphics[width = \linewidth]{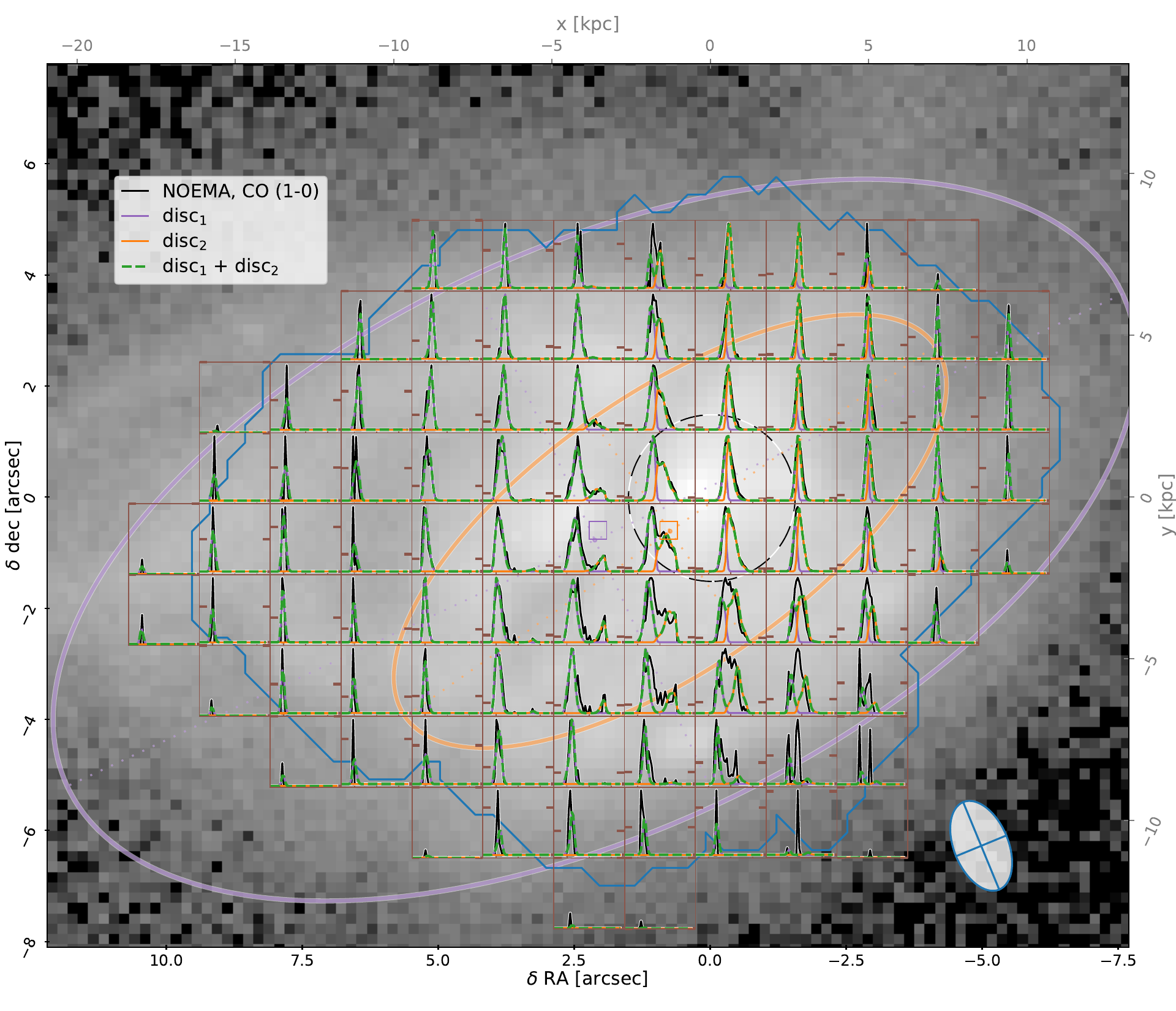}
    \caption{NOEMA CO(1-0) observations and models of two discs, superimposed
on the MegaCam $r$-band image. NOEMA pixels encompassing the centre of the discs are shown. Fits are done
on NOEMA spaxels but for the figure, NOEMA spectra and our fits are binned 16 by 16 and shown in brown boxes
encompassing these binned spaxels. The velocity axis of each box is from -800 to 800 km/s. The two same ticks for values 0,
on the baseline of spectra, and 0.01, are shown on the left and right y axes of each box, to show the scale which is adapted to each
box. The NOEMA elliptical Gaussian synthesised beam is shown on the bottom-right. The outermost contour of NOEMA CO(1-0) intensity is shown in blue.}
    \label{fig:spectra-co-fit}
\end{figure*}

\onecolumn
\section{Two-disc modelling for [OIII]}
\label{app:oiii-two-discs}

\begin{figure*}[h!]
    \centering
    \includegraphics[width = \linewidth]{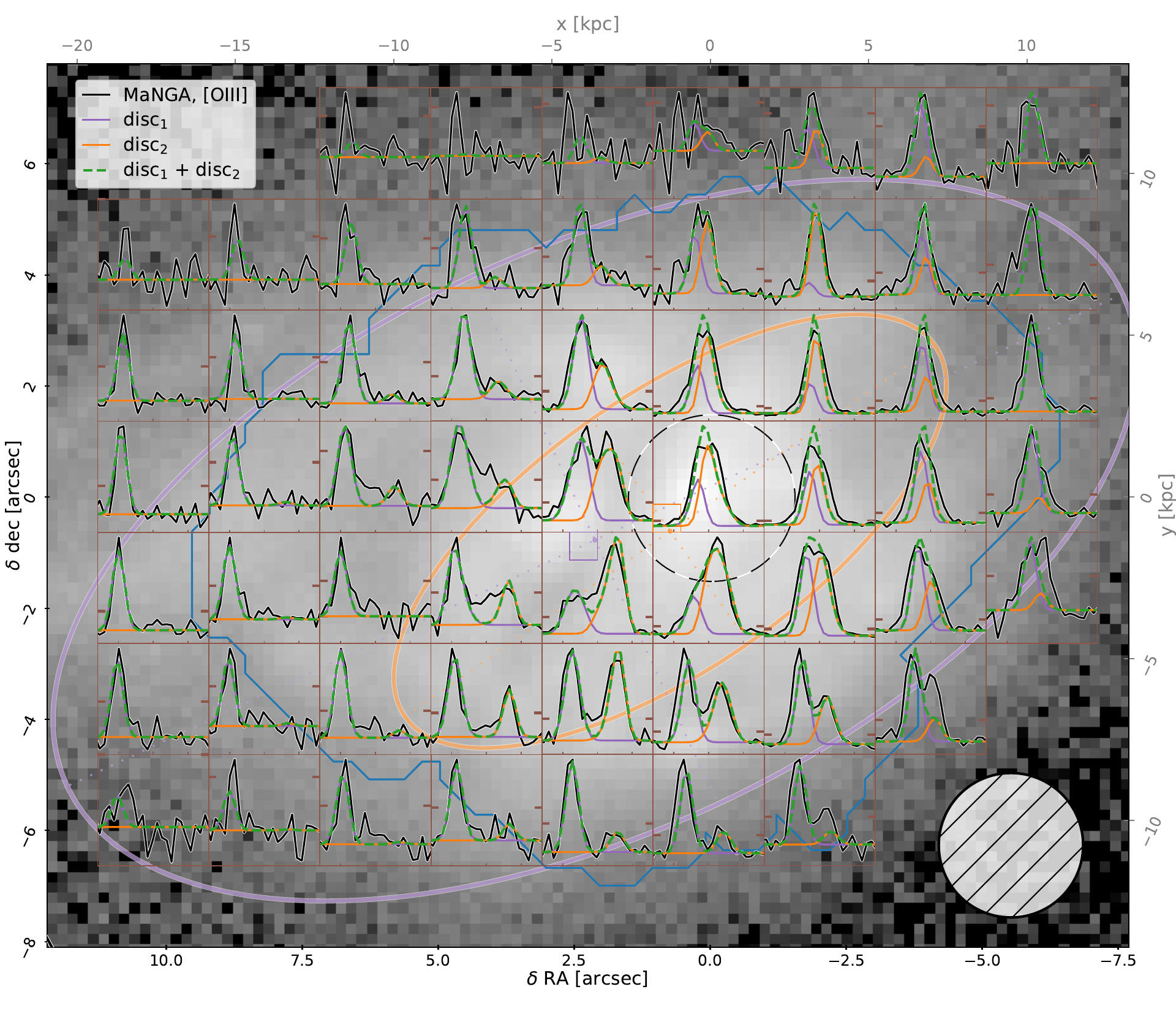}
    \caption{[OIII]~5008 \AA\, MaNGA observations and models of two discs, superimposed
on the MegaCam $r$-band image. MaNGA pixels encompassing the centre of the discs are shown. Fits are done
on original MaNGA spaxels but for the figure, MaNGA spectra and our fits are binned 16 by 16 and shown in brown boxes
encompassing these binned spaxels. The velocity axis of each box is from -800 to 800 km/s. The two same ticks for values 0,
on the baseline of spectra, and 0.01, are shown on the left and right y axes of each box, to show the scale which is adapted to each
box. The Gaussian reconstructed MaNGA PSF is shown on the bottom-right. The outermost contour of NOEMA CO(1-0) intensity is shown in blue.}
    \label{fig:spectra-oiii-fit}
\end{figure*}

\begin{figure*}
    \centering
    \includegraphics[width = \linewidth]{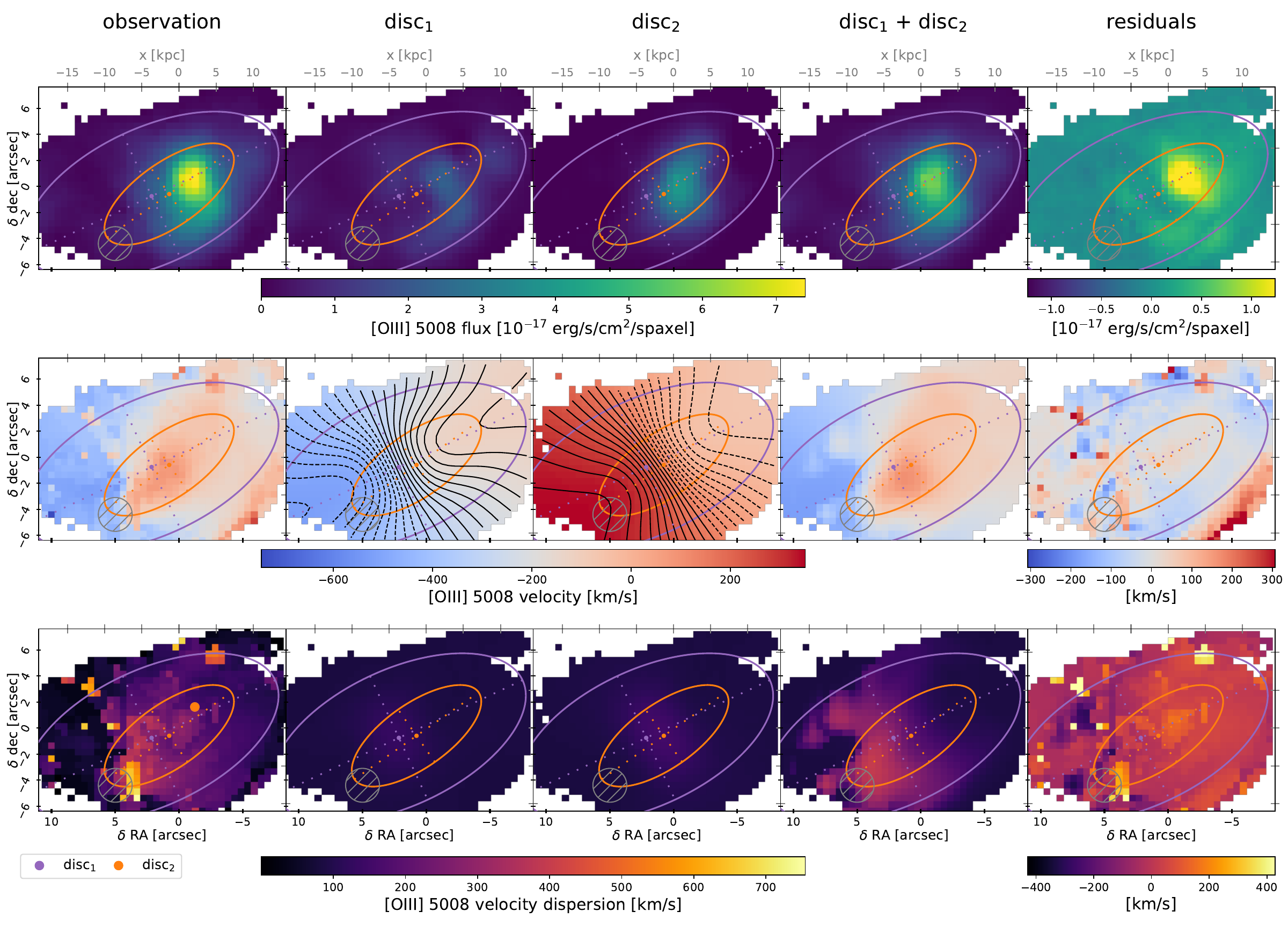}
    \caption{Observations, models and residuals of intensity, velocity and velocity dispersion from the [OIII] 5008 \AA\, line. Models are shown both for each disc (column 2 and 3) and for the two superposed discs (column 4). Residuals are the observations (column 1) minus the two superposed discs (column 4). On the velocity maps of individual discs, iso-line-of-sight velocity curves are represented in black dashed (resp. solid) lines for values below (resp. above) the systemic velocity of the disc.}
    \label{fig:moms-res-oiii-fit}
\end{figure*}

\end{document}